\documentclass[12pt,a4paper,twoside]{article}
\usepackage[dvips]{graphicx}
\usepackage{cite}

\newcommand{\bbbar}  {\ensuremath{\mathrm{b\overline{b}}}}
\newcommand{\epem}   {\ensuremath{\mathrm{e^+e^-}}}
\newcommand{\msb}    {\ensuremath{\mathrm{\overline{MS}}}}
\newcommand{\lmsb}   {\ensuremath{\Lambda_{\mathrm{\overline{MS}}}}}

\newcommand{\as}     {\ensuremath{\alpha_{\mathrm{S}}}}
\newcommand{\ash}    {\ensuremath{\hat{\alpha}_{\mathrm{S}}}}
\newcommand{\ashsq}  {\ensuremath{\hat{\alpha}_{\mathrm{S}}^2}}

\newcommand{\znull}  {\ensuremath{\mathrm{Z^0}}}
\newcommand{\mz}     {\ensuremath{M_{\znull}}}
\newcommand{\asmz}   {\ensuremath{\as(\mz)}}
\newcommand{\mui}    {\ensuremath{\mu_{\mathrm{I}}}}
\newcommand{\mur}    {\ensuremath{\mu_{\mathrm{R}}}}
\newcommand{\azero}     {\ensuremath{\alpha_{\mathrm{0}}}}
\newcommand{\azeromui}  {\ensuremath{\azero(\mui)}}
\newcommand{\azerotwo}  {\ensuremath{\azero({\mathrm{2~GeV}})}}
\newcommand{\oa}     {\ensuremath{\mathcal{O}(\as)}}
\newcommand{\oaa}    {\ensuremath{\mathcal{O}(\as^2)}}

\newcommand{\bt}     {\ensuremath{B_{\mathrm{T}}}}
\newcommand{\bw}     {\ensuremath{B_{\mathrm{W}}}}
\newcommand{\mh}     {\ensuremath{M_{\mathrm{H}}}}
\newcommand{\mhsq}   {\ensuremath{M_{\mathrm{H}}^2}}
\newcommand{\thr}    {\ensuremath{1-T}}
\newcommand{\cp}     {\ensuremath{C}}
\newcommand{\ycut}   {\ensuremath{y_{\mathrm{cut}}}}
\newcommand{\ythree} {\ensuremath{y_3}}
\newcommand{\ca}     {\ensuremath{C_{\mathrm{A}}}}
\newcommand{\cf}     {\ensuremath{C_{\mathrm{F}}}}
\newcommand{\nf}     {\ensuremath{n_{\mathrm{f}}}}
\newcommand{\chisq}  {\ensuremath{\chi^2}}
\newcommand{\chisqd} {\ensuremath{\chi^2/\mathrm{d.o.f.}}}
\newcommand{\xmu}    {\ensuremath{x_{\mu}}}
\newcommand{\eps}    {\ensuremath{\varepsilon}}

\newcommand{\syst}   {\ensuremath{\mathrm{(syst.)}}}

\newcommand{\theo}   {\ensuremath{\mathrm{(theo.)}}}
\newcommand{\fit}    {\ensuremath{\mathrm{(fit)}}}
\newcommand{\lnr}    {\ensuremath{\ln(R)}}
\newcommand{\rs}     {\ensuremath{\sqrt{s}}}

\newcommand{\evis}   {\ensuremath{E_{\mathrm{vis}}}}
\newcommand{\evissq} {\ensuremath{E_{\mathrm{vis}}^2}}

\newcommand{\epsb}   {\ensuremath{\eps_{\mathrm{b}}}}
\newcommand{\bm}[1]  {\mbox{\boldmath\ensuremath{#1}}}
\newcommand{\perc}   {\%}
\newcommand{\momone}[1] {\mbox{\ensuremath{\langle#1\rangle}}}
\newcommand{\sigtot} {\ensuremath{\sigma_{\mathrm{tot}}}}
\newcommand{\rpt}    {\ensuremath{R_{\mathrm{PT}}}}
\newcommand{\rhofit} {\ensuremath{\rho_{\mathrm{fit}}}}
\newcommand{\rhosyst}{\ensuremath{\rho_{\mathrm{syst}}}}
\newcommand{\dd}     {\ensuremath{\mathrm{d}}}
\newcommand{\mil}    {\ensuremath{\mathcal{M}}}
\newcommand{\pp}     {\ensuremath{\mathcal{P}}}

\begin{document}

%
\begin{titlepage}

\pagestyle{empty}

\begin{flushright}
MPI-Ph/2001-005 \\
Revised version, \\
\today
\end{flushright}

\bigskip\bigskip\bigskip\bigskip\bigskip\bigskip\bigskip\bigskip
\begin{center}
  {\huge\bf
    Tests of Power Corrections for \\
    Event Shapes  \\
    in \bm{ \epem } Annihilation
    }
\end{center}
\bigskip\bigskip
\begin{center}
{\Large  P.A.~Movilla~Fern\'andez, S.~Bethke, O.~Biebel, S.~Kluth }
\end{center}

\bigskip

\begin{center}
{\large 
Max-Planck-Institut f\"ur Physik, \\
Werner-Heisenberg-Institut \\
F\"ohringer Ring 6 \\
80805 Munich, Germany \\}
\end{center}

\bigskip

\begin{abstract}
\noindent 
A study of perturbative QCD calculations combined with power
corrections to model hadronisation effects is presented. The QCD
predictions are fitted to differential distributions and mean values of
event shape observables measured in \epem\ annihilation at
centre-of-mass energies from $\rs=14$ to 189~GeV. We investigate the
event shape observables thrust, heavy jet mass, C-parameter, total and
wide jet broadening and differential 2-jet rate and observe a good
description of the data by the QCD predictions. The strong
coupling constant \asmz\ and the free parameter of the power correction
calculations \azerotwo\ are measured to be
\begin{displaymath} 
  \asmz = 0.1171^{+0.0032}_{-0.0020} \;\;\;\mathrm{and}\;\;\;
  \azerotwo = 0.513^{+0.066}_{-0.045} \;\;\;.
\end{displaymath}
The predicted universality of \azero\ is confirmed within the
uncertainties of the measurements. 
\end{abstract}

\bigskip\bigskip\bigskip

\begin{center}
{\Large Acc. by Eur. Phys. J. C }
\end{center}

\end{titlepage}

\section{ Introduction }

The study of hadronic final states in \epem\ annihilation allows
precise tests of the theory of strong interaction, Quantum Chromo
Dynamics (QCD), using event shape observables for the analysis of
hadronic events. For event shape observables perturbative QCD
predictions in \oaa\ and in some cases also in the
next-to-leading-logarithm-approximation (NLLA) are available.  The
various experiments at the PETRA, PEP, TRISTAN, LEP and SLC
colliders collected a large amount of data at centre-of-mass (cms)
energies $\rs=14$ to 189~GeV which can be used to make precise
quantitative tests of QCD. 

Precision tests of perturbative QCD from hadronic event shapes require
a solid understanding of the transition from the perturbatively
accessible partons to the observed hadrons, the hadronisation
process. Hadronisation effects cannot be described directly by
perturbative QCD and are usually estimated by phenomenological
hadronisation models available from Monte Carlo event generators,
e.g. JETSET/PYTHIA~\cite{jetset3}, HERWIG~\cite{herwig} or
ARIADNE~\cite{ariadne3}.

Alternatively, analytical approaches are pursued in order to deduce as
much information as possible about hadronisation from the perturbative
theory. Hadronisation contributions to event shape observables
evolve like reciprocal powers of the hard interaction scale
\rs\ (power corrections)~\cite{webber94,dokshitzer95}. An analytic
model by Dokshitzer, Marchesini and Webber (DMW) of hadronisation
valid for some event shape observables derives the structure of the
power corrections from perturbative QCD. The model assumes that the
strong coupling remains finite at low energy scales where simple
perturbative calculations break
down~\cite{dokshitzer95a,dokshitzer97a}. The model parametrises the
magnitude of non-perturbative effects by introducing moments of the
running strong coupling \as\ as parameters to be determined by
experiment.

Several experimental tests of power corrections in the DMW model with
differential distributions or 1st moments (mean values) of event shape
observables measured in \epem\ annihilation have been
done~\cite{jadenewas,jadec,delphias133,l3lep2data1,delphilep2data1,dokshitzer97a,dokshitzer99a,biebel98,movilla98,wicke98}.
In the present paper we test power corrections in the DMW model to the
differential distributions and mean values of event shape observables
measured in \epem\ annihilation experiments at $\rs=14$ to 189~GeV.

We use resummed \oaa+NLLA QCD calculations combined with power
corrections to fit the event shape distributions with \asmz\ and the
non-perturbative parameter \azero\ as free parameters. In the case of
the mean values \oaa\ calculations together with power corrections are
fitted to the data. We investigate the prediction of the DMW model
that the non-perturbative parameter does not depend on the specific
event shape observable, i.e. that it is universal.

Section~\ref{sec_theory} starts with an overview of the observables and
briefly explains the theoretical predictions. The data used in our
study and the fit results are presented in section~\ref{sec_anal}. In
section~\ref{sec_conc} we give a summary and draw conclusions from our
results.

\section{ QCD Predictions }
\label{sec_theory}

\subsection{ Event Shape Observables }

We employ the differential distributions and mean values of the event
shape observables thrust, heavy jet mass, C-parameter, total and wide
jet broadening. The mean value of the differential 2-jet rate based on
the Durham algorithm is used as well. The definitions of these
observables are given in the following:
\begin{description}

\item[Thrust \bm{T}]
 The thrust value is given by the expression~\cite{thrust1,thrust2}
\begin{displaymath} 
  T = \max_{\vec{n}}\left( \frac{\sum_i |
     \vec{p}_i \cdot \vec{n} | } {\sum_i | \vec{p}_i | } \right)\ \ .
\end{displaymath}
where $\vec{p}_i$ are the momentum vectors of the particles in an
event. The thrust axis $\vec{n}_T$ is the vector $\vec{n}$ which
maximises the expression in parentheses. We use \thr\ in this
analysis, because in this from the distribution is comparable to
those of the other observables. A plane perpendicular to $\vec{n}_T$
through the origin divides the event into two hemispheres $H_1$ and
$H_2$ which are used in the defintion of heavy jet mass and the jet
broadening observables below. 

\item[Heavy Jet Mass \bm{\mh}]
The invariant mass $M_i$ of all particles contained in hemisphere
$H_1$ or $H_2$ is calculated~\cite{jetmass}. The observable \mh\ is
defined by
\begin{displaymath} 
  \mh= \max(M_1,M_2)/\rs\;\;\;.  
\end{displaymath}
Some experiments use the definition $\mhsq=\max(M_1,M_2)^2/s$ where in
\oa\ we have the relation $\thr=\mhsq$.

\item[Jet Broadening]
The jet broadening measures are calculated by~\cite{nllabtbw}:
\begin{displaymath}
   B_k = \frac{\sum_{i\in H_k} | \vec{p}_i
      \times \vec{n}_T | } {2\sum_i | \vec{p}_i | } 
\end{displaymath}
for each hemisphere $H_k$, $k=1,2$. The
total jet broadening is given $\bt=B_1 + B_2$ and the wide jet
broadening is defined by $\bw=\max(B_1,B_2)$.

\item[C-parameter]
The $C$-parameter is defined as~\cite{parisi78,donoghue79}
\begin{displaymath}
   C = 3 (\lambda_1 \lambda_2 + \lambda_2 \lambda_3 +
        \lambda_3 \lambda_1 ) 
\end{displaymath} 
where $\lambda_k$, $k=1,2,3$, are the eigenvalues of the
momentum tensor 
\begin{displaymath} 
  \Theta^{\alpha\beta} = \frac{\sum_i
  (p_i^{\alpha}p_i^{\beta}) / |\vec{p}_i|} {\sum_i
  |\vec{p}_i|} \;\;,\;\;\; \alpha,\beta=1,2,3 \;\;\;.
\end{displaymath}

\item[Differential 2-jet rate]
The differential 2-jet rate is determined using the Durham jet finding
algorithm~\cite{durham}. In this algorithm the quantity
$y_{ij}=2\min(E_i^2,E_j^2)(1-\cos\theta_{ij})/\evissq$, $\evis=\sum_k
E_k$, is computed for all pairs of (pseudo-) particles with energies
$E_i$, $E_j$ in the event. The pair with the smallest $y_{ij}$ is
combined into a pseudo particle by adding the 4-vectors and the
procedure is repeated until all $y_{ij}>\ycut$.  The value of \ycut\
where the number of jets in an event changes from three to two is
called \ythree. The differential 2-jet rate is defined by the
differential distribution of \ythree.

\end{description}

\subsection{ Perturbative QCD predictions }

We use QCD predictions in \oaa\ matched with resummed NLLA
calculations in our analysis of differential
distributions~\cite{nllathmh,nllabtbw2,nllacp}. In the 
NLLA the cumulative distribution $R(y)=\int_0^y
1/\sigtot(\dd\sigma/\dd y')\dd y'$ of an observable $y$ is considered. 
The NLLA is valid in regions of phase space where $y$ is small,
i.e. where the emission of multiple soft gluons from a system of
approximately back-to-back and hard quarks dominates (2-jet
region). QCD predictions in \oaa\ are expected to be valid in regions
of phase space where emmission of a single hard gluon dominates (3-jet
region). We choose to combine the \oaa\ with the NLLA calculations
with the \lnr-matching scheme, because it has theoretical
advantages~\cite{nllathmh,dokshitzer97a} and is also preferred in
experimental
analysis~\cite{OPALPR075,OPALPR134,OPALPR158,OPALPR197,OPALPR299,OPALPR303,jadenewas,colrun}.
Other matching schemes exist and will be considered in the study of
systematic uncertainties, see section~\ref{sec_syst} below. 

The complete perturbative QCD prediction renormalised at the scale
\mur\ for a cumulative distribution $\rpt(y)$ using the \lnr-matching
scheme takes the following form~\cite{nllathmh,OPALPR075}:  
\begin{eqnarray}
\label{equ_lnr}
  \ln\rpt(y) & = & Lg_1(L\ash(\mur))+g_2(L\ash(\mur)) \\ \nonumber
             &   & - (G_{11}L+G_{12}L^2)\ash(\mur)
                   - (G_{22}L^2+G_{23}L^3)\ashsq(\mur) \\ \nonumber
             &   & + A(y)\ash(\mur)
   + \left( B(y)-2A(y)-\frac{1}{2}A(y)^2 \right)\ashsq(\mur)\;\;\;, \\
  \nonumber
\end{eqnarray}
where $L=\ln(1/y)$ and $\ash=\as/(2\pi)$. The functions $g_1$ and
$g_2$ represent the all-orders resummations of leading and subleading
logarithmic terms, respectively, and the $G_{nm}$ coefficients are
given e.g. in~\cite{OPALPR075}. The coefficient functions $A(y)$ and
$B(y)$ are defined by $A(y)=\int_0^y(\dd A/\dd y')\dd y'$ and
$B(y)=\int_0^y(\dd B/\dd y')\dd y'$, respectively. The differential
distributions $\dd A/\dd y$ and $\dd B/\dd y$ are obtained by
integration of the \oaa\ QCD matrix elements using the program
EVENT2~\cite{event2}. The prediction is normalised to the total
hadronic cross section evaluated in \oa.

The renormalisation scale \mur\ is identified with the cms energy
$\rs=Q$ of the measurement. The dependence of the perturbative QCD
predictions on the renormalisation scale is studied by introducing the
renormalisation scale parameter $\xmu=\mur/Q$ and making the
replacements of~\cite{OPALPR075}, equation (23). 

The mean values of event shape observable distributions are defined by
\begin{equation}
  \momone{y} =
  \int_0^{y_{max}}y\frac{1}{\sigtot}\frac{\dd\sigma}{\dd y}\dd y\;\;\;,
\end{equation}
where $y_{max}$ is the largest possible value of the observable $y$
(kinematic limit). The perturbative QCD prediction of mean values
$\momone{y}_{\mathrm{PT}}$ in \oaa\ is given by 
\begin{equation}
  \momone{y}_{\mathrm{PT}} = {\cal A}_y \ash(\mur) + 
  \left( {\cal B}_y + 
    (\pi\beta_0\ln(\xmu^2)-1)2 {\cal A}_y \right)\ashsq(\mur)
\label{equ_oaa}
\end{equation}
where $\beta_0=(33-2\nf)/(12\pi)$ with the number of active quark
flavours $\nf=5$ at the cms energies considered here. The \oa\ and
\oaa\ coefficients ${\cal A}_y$ and ${\cal B}_y$ are taken
from~\cite{biebel01a}. Calculations of mean values in NLLA are not yet
available, because the NLLA predictions diverge for very small values
of $y$ and do not vanish at the kinematic limits $y_{max}$ of the
observables.

\subsection{ Power Corrections }
\label{sec_powcor}

Non-perturbative effects to event shape observables are calculated in
the DMW model as contributions from gluon radiation at low energy
scales where perturbative evolution of the strong coupling breaks
down. The location of the divergence in the perturbative evolution of
\as, known as the Landau pole, is given by $\lmsb\simeq200$~MeV in the
\msb\ renormalisation scheme. The model assumes that the physical
strong coupling remains finite at scales around and below the Landau
pole. A new free non-perturbative parameter 
\begin{equation}
  \azeromui = \frac{1}{\mui} \int_0^{\mui} \as(k) {\mathrm{d}}k
\end{equation}
is introduced to parametrise the unknown behaviour of
$\as(Q)$ below the so-called infrared matching scale \mui. The
non-perturbative and the perturbative evolution of the strong coupling
are merged at the scale \mui\ which is generally taken to be
2~GeV~\cite{dokshitzer98b}. 

The power corrections are calculated including two loop corrections
for the differential distributions of the event shape observables
considered here~\cite{dokshitzer98b,dokshitzer99a}. The effect of
hadronisation on the distribution obtained from experimental data is
described by a shift of the perturbative prediction away from the
2-jet region:
\begin{equation}
  \frac{d\sigma}{dy} = \frac{d\sigma_{\mathrm{PT}}}{dy}(y-\pp D_y)\;\;\;,
\label{equ_pc}
\end{equation}
where $y=\thr$, \mhsq, \cp, \bt\ and \bw. The
factor \pp\ depends on non-perturbative parameter \azero\ and is
predicted to be universal~\cite{dokshitzer98b}: 
\begin{equation}
  \pp = \frac{4\cf}{\pi^2} \mil \frac{\mui}{Q}
  \left( \azeromui - \alpha_s(\mur) -\beta_0\frac{\alpha_s^2(\mur)}{2\pi}
    \left( \ln\frac{\mur}{\mui}+\frac{K}{\beta_0}+1 \right) \right)
\label{equ_pcp}
\end{equation}
with the colour factor $\cf=4/3$. The factor
$K=(67/18-\pi^2/6)\ca-(5/9)\nf$ originates from the choice of the
\msb\ renormalisation scheme. The Milan factor \mil\ accounts for
two-loop effects and its numerical value is
1.49~\cite{dokshitzer99b}. The theoretical uncertainty of \mil\ is about
20\% due to missing higher order corrections~\cite{dokshitzer98b}. The
quantity $D_y$ depends on the
observable~\cite{dokshitzer98b,dokshitzer99a}: 
\begin{equation}
  D_{\thr}=2\;,\;\;D_{\mhsq}=1\;,\;\;D_{\cp}=3\pi\;,
\label{equ_pcd}
\end{equation}
\begin{displaymath}
  D_b = a_b\ln \frac{1}{b} + 
                  F_b(b,\as(bQ))\;,\;\; b=\bt,\bw \;,\;\;
                  a_{\bt}=1, a_{\bw}=\frac{1}{2}\;.
\end{displaymath}
A simple shift is expected for \thr, \mhsq\ and \cp\ whereas for the
jet broadening variables \bt\ and \bw\ an additional
squeeze\footnote{The term ``squeeze'' refers to the form of event
shape distributions which are more peaked in perturbative predictions
compared to the predictions including hadronisation effects.} of the
distribution is predicted. The more complex behaviour for the jet
broadening observables calculated in~\cite{dokshitzer99a} is related
to the interdependence of non-perturbative and perturbative effects
which cannot be neglected for these observables. The power corrections
for \bt\ and \bw\ predict in addition to the shift an increasing
squeeze with decreasing cms energies. The necessity of an additional
non-perturbative squeeze of the jet broadening distributions was
already pointed out in~\cite{movilla98}.

The power corrections for mean values of \thr, \mhsq\ and \cp\  are
obtained by taking the first moment of equation~(\ref{equ_pc}) and read:
\begin{equation}
  \momone{y} = \momone{y}_{\mathrm{PT}} + \pp D_y\;\;\;.
\label{equ_pcmom}
\end{equation}
In the case of mean values of \bt\ and \bw\ the predictions
from~\cite{dokshitzer99a} are used.  For the observable \ythree\ the
leading power correction is expected to be of the type $1/Q^2$ or
$(\ln Q)/Q^2$~\cite{dokshitzer95a} but the corresponding coefficients
are not yet calculated.

\section{ Analysis of the Data }
\label{sec_anal}

\subsection{ Data Sets }

Experimental data below the \znull\ peak are provided by
the experiments of the PETRA (12 to 47~GeV, about 50000 events in
total), PEP (29~GeV, about 28000 events in total) and TRISTAN 
(55 to 58~GeV, about 1200 events in total) colliders. Data around the
\znull\ resonance are from the four LEP experiments with
$\mathcal{O}(10^5)$ events per experiment and from SLD (about 40000
events) while data above the \znull\ are exclusively from the LEP
experiments with $\mathcal{O}(10^2)$ events per experiment from
$\rs=133$ to 183~GeV and $\mathcal{O}(10^3)$ per experiment at
$\rs=189$~GeV.  For event shape distributions table~\ref{tab_data}
gives the references and also the ranges considered in the fits (see
section~\ref{sec_fits} below). For mean values we consider published
data available in the energy range of 13 up to 189~GeV
\cite{alephlep1data,alephas133,
amydata,    
delcodata,  
delphilep1data, delphias133, delphilep2data1, 
jadenewas,
hrsdata,
l3lep1data, l3isrfsr2, l3lep2data1, l3as133, l3115, l3162,
markiidata,
markjdata1, markjdata2,
OPALPR054, OPALPR158, OPALPR197, OPALPR303,
sldnlla,
tassodata2, tassodata}.
All data used in this study are corrected for the limited resolution
and acceptance of the detectors and event selection criteria and are
published with statistical and experimental systematic uncertainties.

\subsection{ Fit Procedure }
\label{sec_fits}

The standard analyses use the entire data sets as described above. For
the perturbative predictions of differential distributions we employ
the matched resummed \oaa+NLLA QCD prediction given by
equation~(\ref{equ_lnr}) while the power corrections are implemented
according to equation~(\ref{equ_pc}). For mean values the \oaa\
perturbative prediction from equation~(\ref{equ_oaa}) is used combined
with the power corrections according to equation~(\ref{equ_pcmom}).

For each observable we perform simultaneous \chisq-fits 
with \asmz\ and \azerotwo\ as free parameters. The strong coupling
\asmz\ is evolved to the renormalisation scale $\mur=\xmu Q$ with
$Q=\rs$ of a given event shape distribution or mean value using the
two-loop formula for the running coupling~\cite{ellis96}. The \chisq\
is defined by $\chisq = \sum_i ((d_i-t_i)/\sigma_i)^2$ where $d_i$ is
the value of measurement $i$, $t_i$ is the corresponding theoretical
prediction and $\sigma_i$ is the quadratic sum of statistical and
experimental systematic uncertainties of $d_i$.

\begin{figure}[!htb]
\begin{center}
\begin{tabular}{cc}
\includegraphics[width=0.475\textwidth]{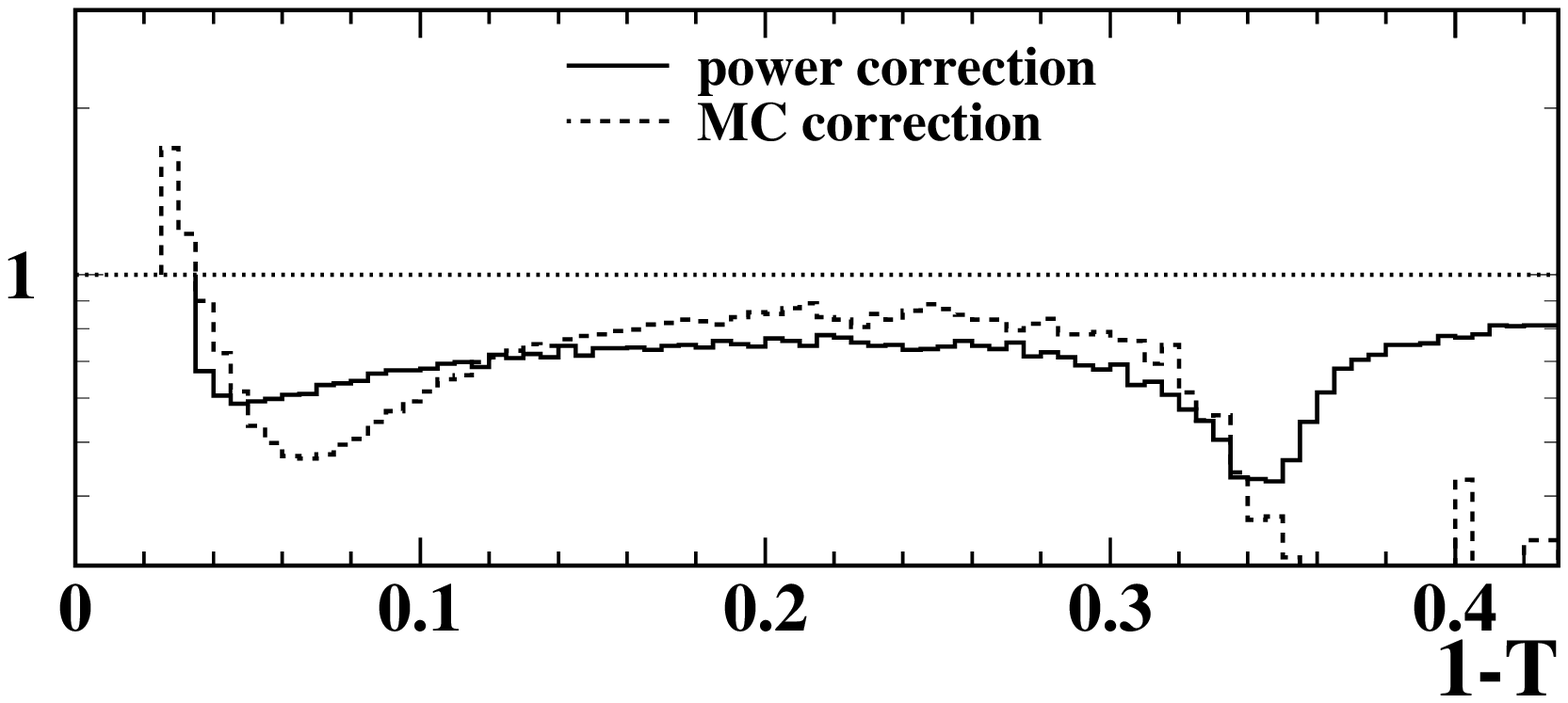} &
\includegraphics[width=0.475\textwidth]{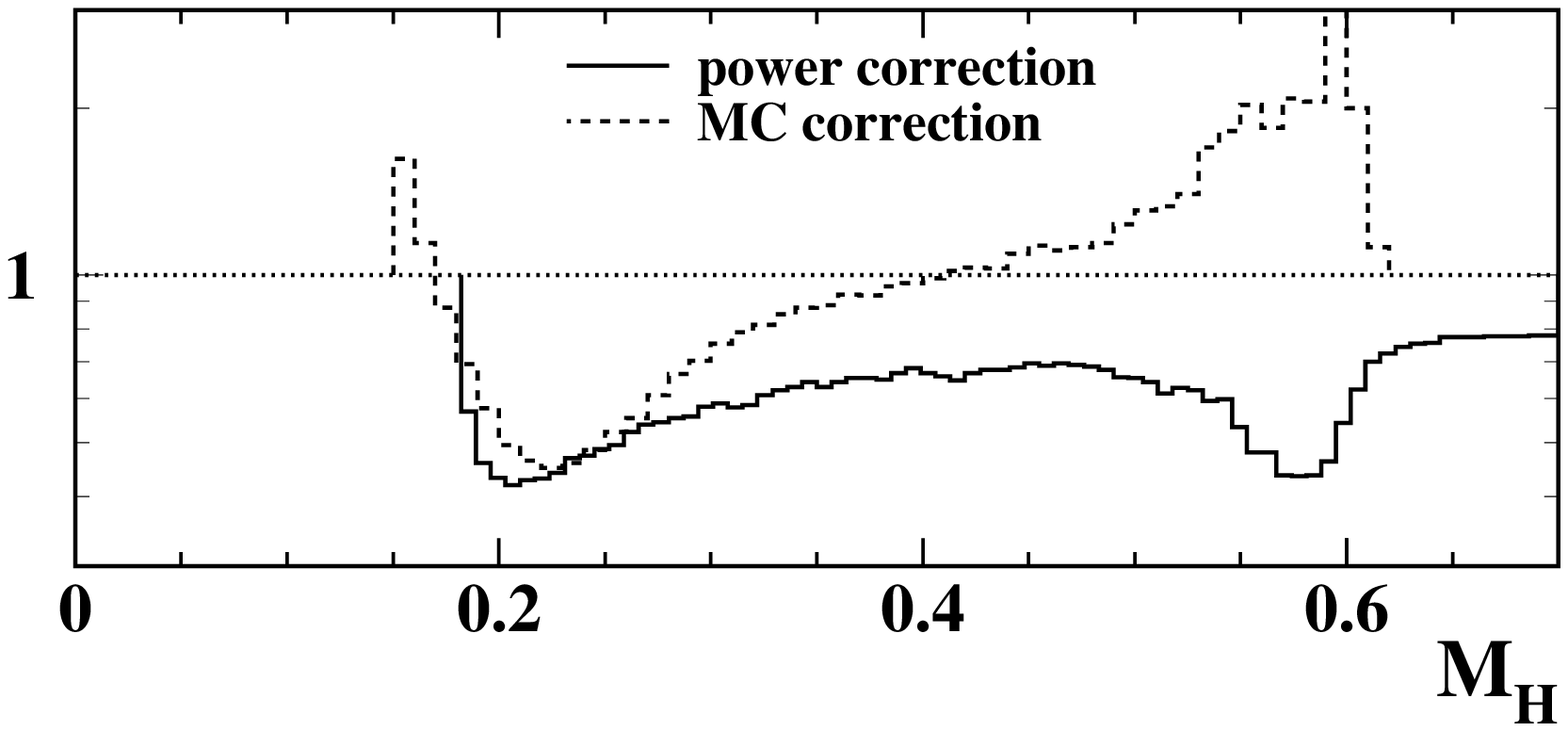} \\
\includegraphics[width=0.475\textwidth]{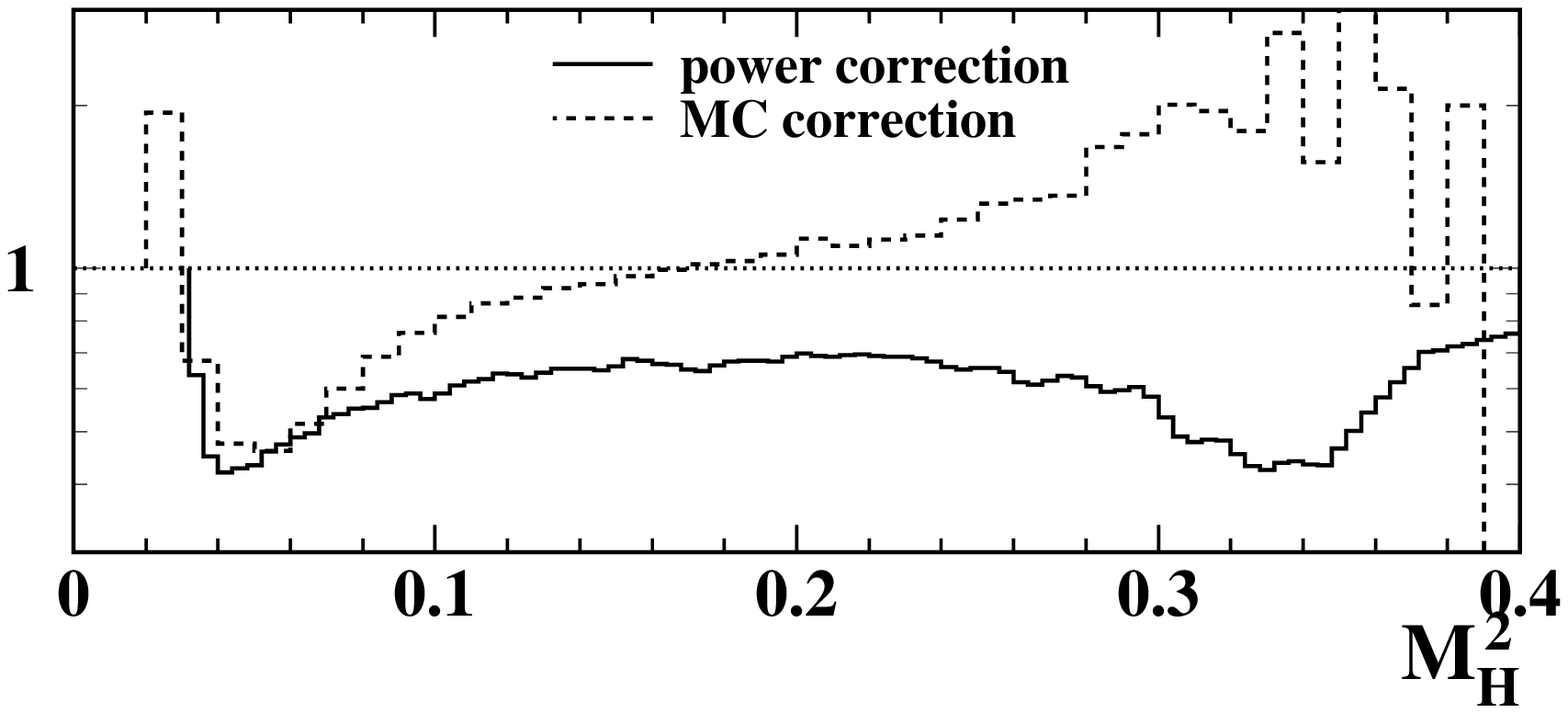} &
\includegraphics[width=0.475\textwidth]{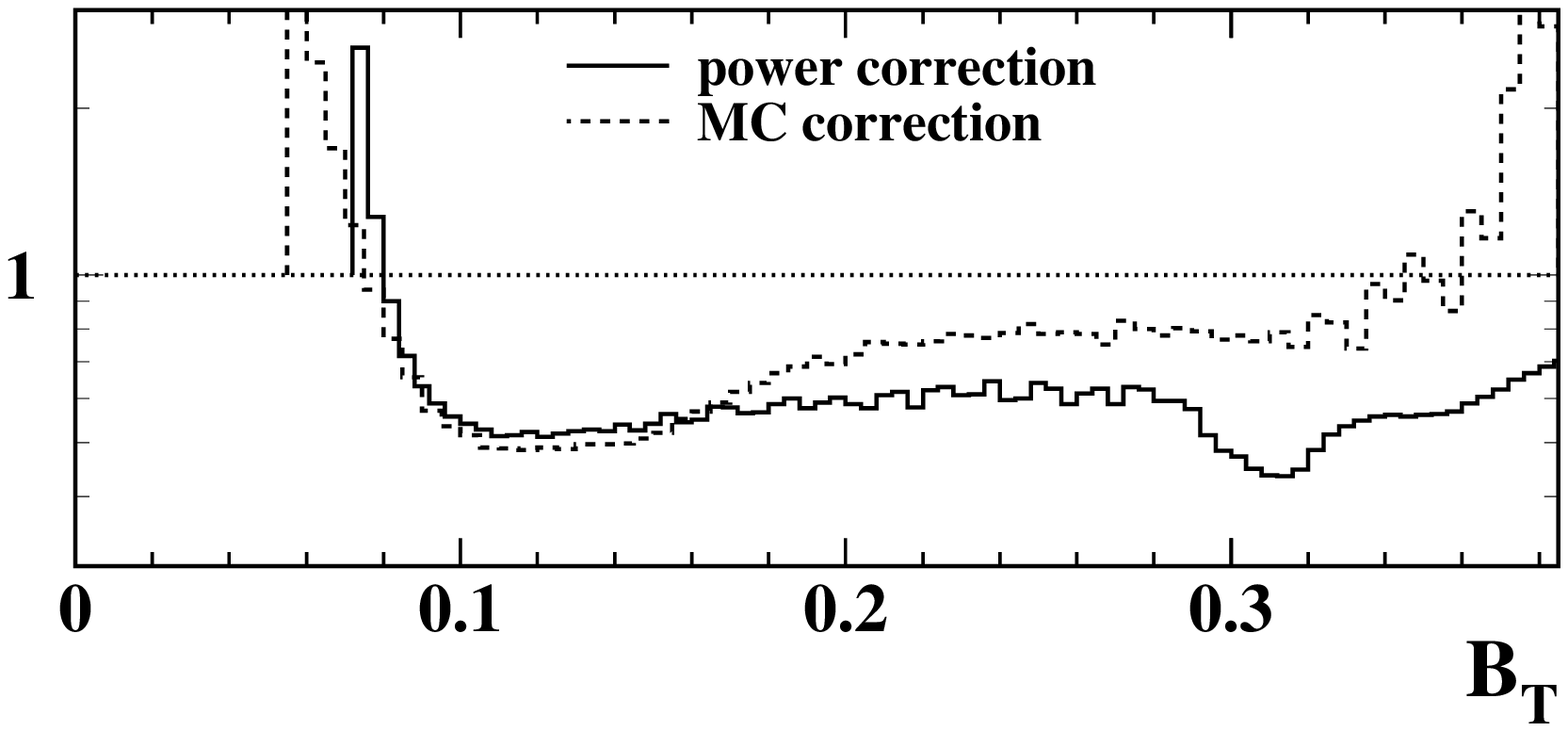} \\
\includegraphics[width=0.475\textwidth]{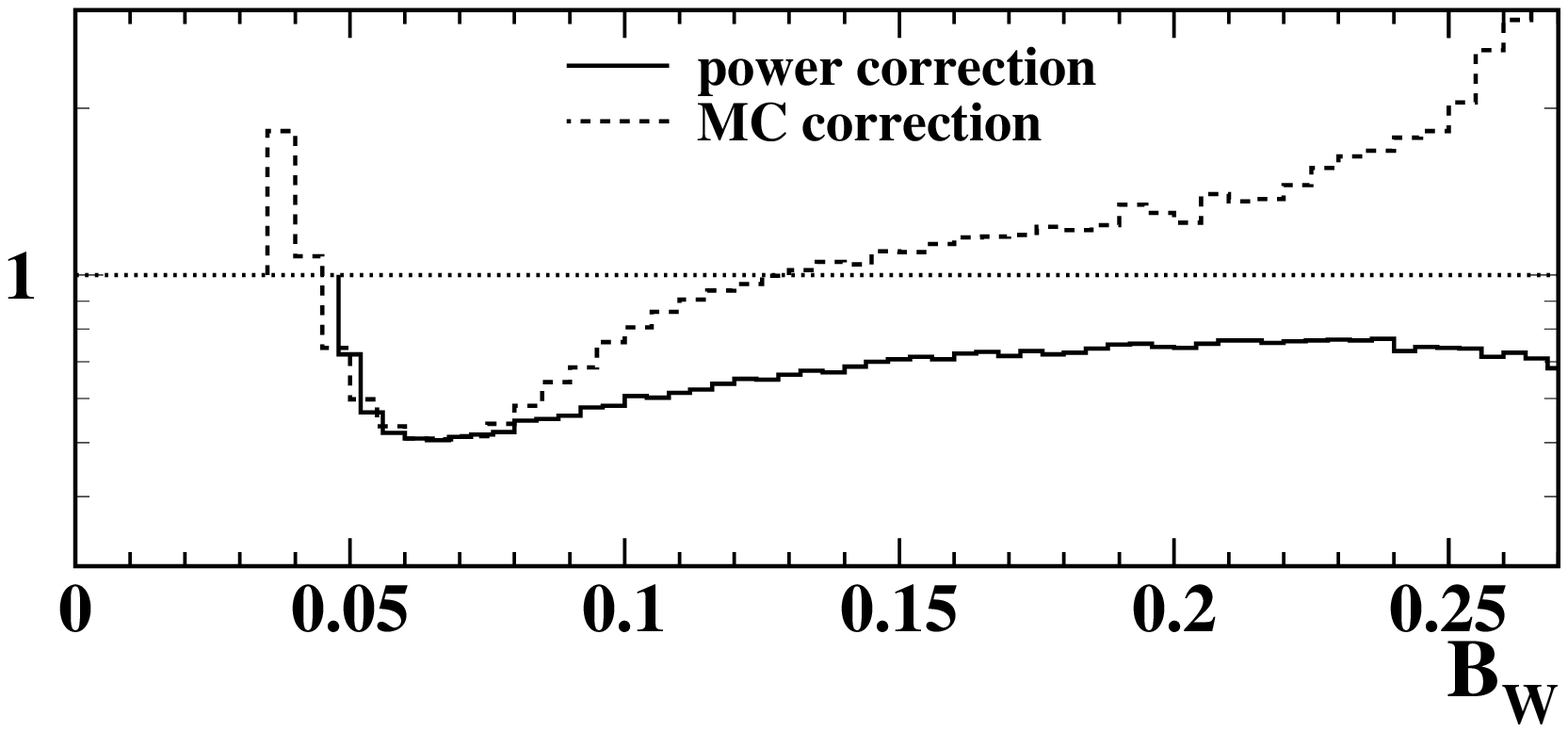} &
\includegraphics[width=0.475\textwidth]{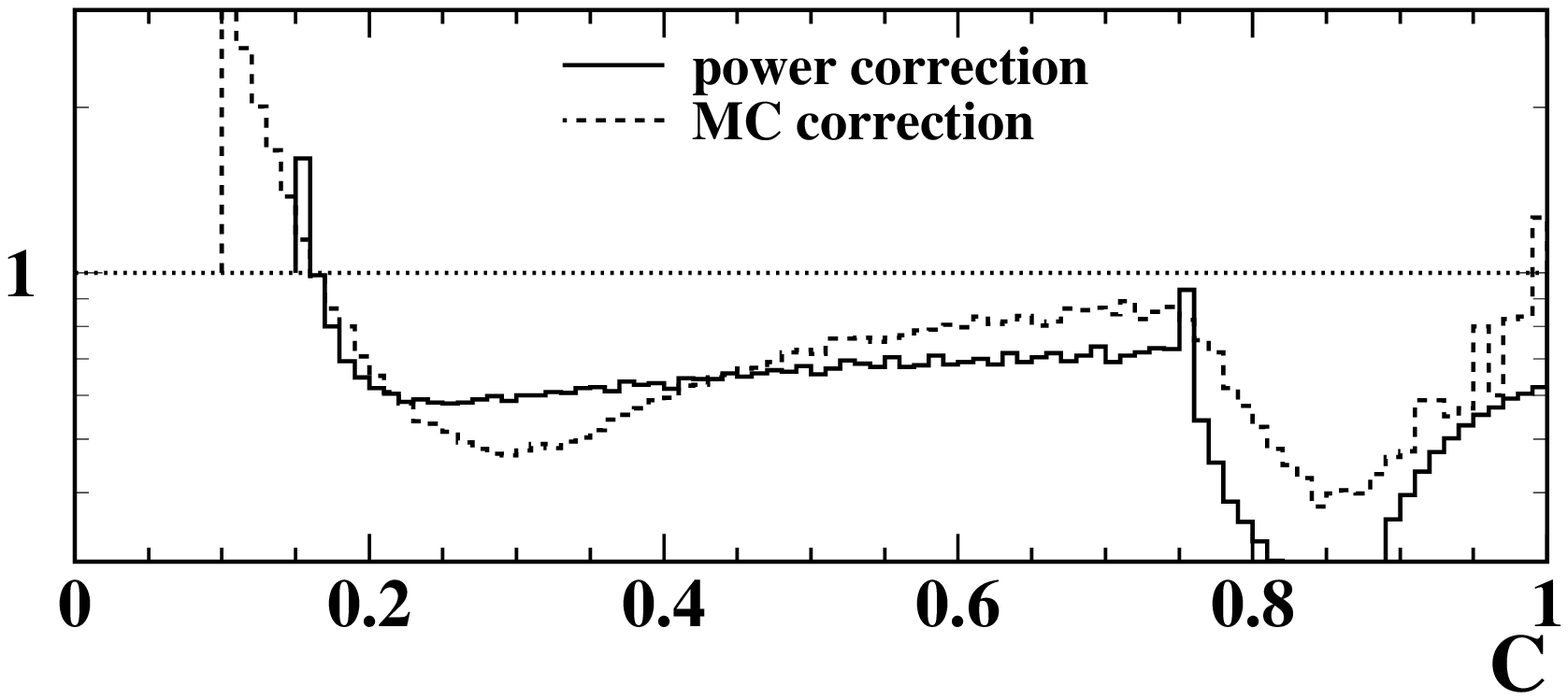} \\
\end{tabular}
\caption[bla]{ The figure presents hadronisation correction factors at
$\rs=35$~GeV
estimated using power corrections (solid lines) or using the JETSET
Monte Carlo program (dashed lines). The hadronisation corrections for
power corrections are given by the ratio of the perturbative QCD
prediction over the same prediction combined with power corrections
using the fitted values of \asmz\ and \azerotwo. The Monte Carlo
hadronisation corrections are given by the ratio of distributions
calculated at the parton- and hadron-level, respectively. }
\label{fig_pcr}
\end{center}
\end{figure}

The fit ranges for fits of event shape distibutions are defined
individually for each cms energy such that the 2-jet region of the
distribution is exploited as far as possible.  The fit ranges are
limited by the demands i) that the \chisq\ of the extreme bins do not
contribute substantially to the total \chisq\ of the distribution, ii)
that the perturbative QCD prediction is reliable, and iii) that the
power corrections are under control. Requirement iii) is checked by
monitoring the ratio of the theoretical predictions without and with
power corrections, respectively, using the fit results for \asmz\ and
\azerotwo. Figure~\ref{fig_pcr} (solid lines) presents these ratios for
\thr, \mh, \mhsq, \bt, \bw\ and \cp\ at $\rs=35$~GeV. The fit ranges
are chosen such that regions of rapidly varying power corrections are 
excluded. The chosen fit ranges are listed in table~\ref{tab_data}.

\subsection{ Effects of \bm{ \bbbar } Events at low \bm{ \rs } }
\label{sec_bcorr}

The presence of events from the reaction $\epem\rightarrow\bbbar$ at
low cms energies \rs\ can distort the event shape distributions,
because the effects of weak decays of heavy B-hadrons on the topology
of hadronic events cannot anymore be neglected. An additional
 problem arises from comparing QCD calculations based on
massless quarks with data containing massive quarks at \rs\ close
to the production threshold.

\begin{figure}[!htb]
\begin{center}
\includegraphics[width=0.6\textwidth]{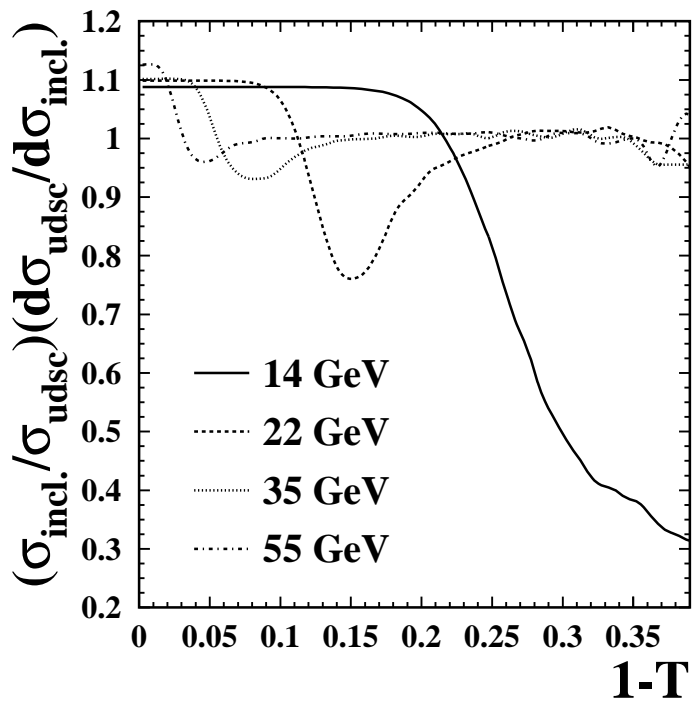}
\caption[bla]{ The figure presents ratios of distributions of \thr\
calculated using u, d, s, and c quarks events or all events using
Monte Carlo simulation. The different line types indicate the cms
energy at which the Monte Carlo simulation was run. }
\label{fig_bcorr}
\end{center}
\end{figure}

At $\rs\ll\mz$ \bbbar\ events constitute about 9\% of the total
event samples. Ideally one would correct the data experimentally by
identifying \bbbar\ events and removing them from the sample. However,
since we have only published event shape data without information on
specific quark flavours we resort to a correction based on Monte
Carlo simulations. We generate samples of $10^6$ events at each
\rs\ with the JETSET~7.4 program~\cite{jetset3} with the parameter
set given in~\cite{OPALPR141}. For each event shape observable we
build the ratio of distributions calculated with u, d, s and c quark
events to those calculated with all events. This ratio is multiplied
with the bin contents of the data to obtain corrected
distributions. This procedure is applied to all data at $\rs<\mz$. We
correct the mean values for the contribution from b quarks using
exactly the same procedure as for the correction of the differential
distributions. It was verified that the simulation provides an
adequate description of the data at all values of
$\rs<\mz$. Figure~\ref{fig_bcorr} shows the ratio of distributions of
\thr\ calculated using u, d, s, and c quark events or all events
obtained at $\rs=14$ to 55 GeV as an example. The correction is
reasonable within the fit ranges at $\rs>14$~GeV while at $\rs=14$~GeV
the correction is a large effect.

Systematic effects due to uncertainties in the Monte Carlo parameters
are expected to be small for the ratio except for those parameters
which only affect the \bbbar\ events in the samples. The most
important such parameter is the value of \epsb\ in the Peterson
fragmentation function~\cite{peterson83} which controls the
fragmentation of b quarks in the simulation. Threshold effects on the
fraction of \bbbar\ events at low \rs\ which depend on the value of
the b-quark mass in the simulation are found to be negligible for the
fit results.

\subsection{ Fit and Systematic Uncertainties }
\label{sec_syst}

We consider the following for both fits to differential distributions
and to mean values unless specified otherwise:
\begin{description}

\item[Fit error]
The fit errors for \asmz\ and \azerotwo\ are taken from the diagonal
elements of the error matrix after the fit has converged.

\item[Renormalisation scale]
Systematic uncertainties from perturbation theory are assessed by
varying \xmu\ between 0.5 and 2.0. The changes in the fit results
w.r.t. the standard results are taken as asymmetric systematic
uncertainties. In the case that both deviations have the same sign the
larger one defines a symmetric uncertainty. 

\item[Matching scheme]
As a further systematic check in the analysis of differential
distributions we use different matching schemes, namely the modified
\lnr- and $R$-matching schemes, to combine
the \oaa\ with the resummed NLLA calculations~\cite{OPALPR075}. A
possible matching scheme uncertainty is defined by the larger
deviation caused by using \lnr- or $R$-matching. 

\item[Power corrections]
Uncertainties due to the power corrections come from the choice of
the value of \mui\ and from the theoretical uncertainty of the Milan
factor \mil. We vary \mui\ by $\pm 1$~GeV and \mil\ by $\pm 20\%$ and
take in both cases changes of the fit results w.r.t. the standard
results as asymmetric systematic uncertainties. No error contribution
from the variation of \mui\ is assigned to \azeromui, because setting
\mui\ to a different value corresponds to a redefinition of \azeromui.

\item[Fragmentation of b quarks]
The standard analysis is carried out with corrected data at
$\rs<\mz$ based on the JETSET tuning of~\cite{OPALPR141} as
explained in section~\ref{sec_bcorr}. The value of the JETSET
parameter \epsb\ is varied around its central value
$\epsb=0.0038\pm0.0010$ by adding or subtracting its error and the
analysis including correction of the data at $\rs<\mz$ is
repeated. Deviations w.r.t. the standard results are considered as
asymmetric uncertainties.

\item[Experimental uncertainties]
We examine the dependence of the results on the input data taken for
the fits in several ways:
\begin{enumerate}
\item
We perform the fit without the LEP/SLC data at $\rs\simeq\mz$.
\item
In the case of fits to distributions the fits are repeated using
seperately either the data below or above the \znull\ peak. This also
checks for possible higher order non-perturbative contributions to the
power corrections. Such tests are impractical with mean values,
because the sensitivity to the power corrections is reduced
when only restricted ranges in \rs\ are used, in particular for
$\rs>\mz$.  
\item
A further source of systematic uncertainty in the analysis of
differential distributions only comes from the choice of the fit
ranges. The lower and the upper edges of the fit ranges of all
distributions of a given observable are varied in both directions by
one bin. We take the largest of the four deviations w.r.t. the standard
result as a systematic uncertainty.
\end{enumerate}
For distributions the largest deviation from 1. and 2. w.r.t. the
standard results is added in quadrature with the uncertainty from
3. and the result defines a symmetric systematic uncertainty. 
For mean values only the deviation from 1. is taken to define a
symmetric systematic uncertainty.

\end{description}

The total errors of the standard results are defined as the quadratic
sum of the fit errors, the renormalisation scale uncertainty, the
power correction and the experimental uncertainties. In the case of
distributions the larger of the matching scheme and the
renormalisation scale uncertainties is included in the total error and
the fit range uncertainty is added as well.

\subsection{ Results of Fits to Event Shape Distributions }

\begin{figure}[!htb]
\begin{center}
\includegraphics[width=0.7\textwidth]{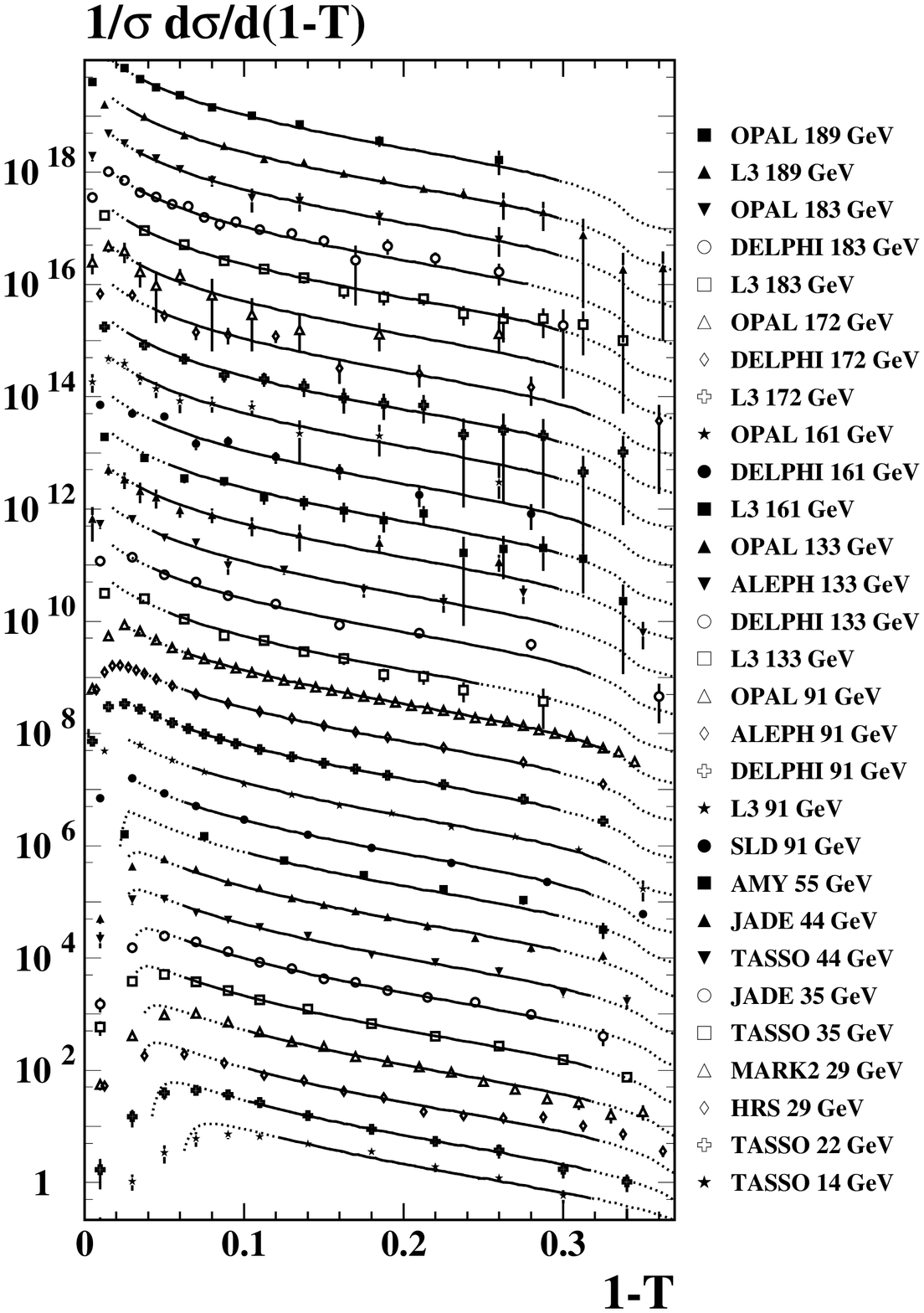}
\caption[bla]{ Scaled distributions for \thr\ measured at
$\rs=14$ to 189~GeV. The error bars indicate the total errors of the
data points. The solid lines show the result of the simultaneous fit
of \asmz\ and \azero\ using resummed \oaa+NLLA QCD predictions with
the ln(R)-matching combined with power corrections. The dotted lines
represent an extrapolation of the fit result. }
\label{fig-t_plot}
\end{center}
\end{figure}

\begin{figure}[!htb]
\begin{center}
\includegraphics[width=0.7\textwidth]{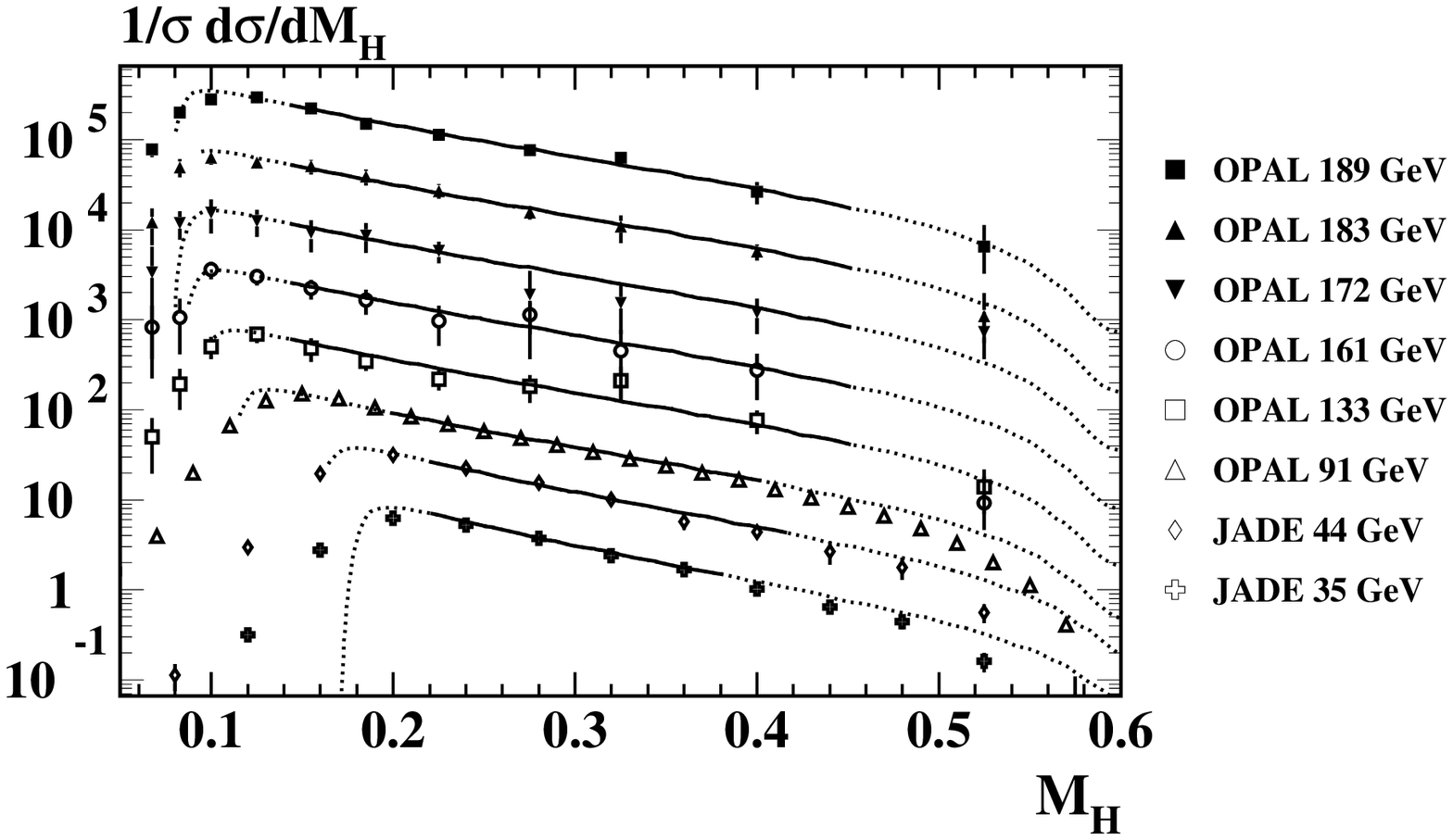}
\includegraphics[width=0.7\textwidth]{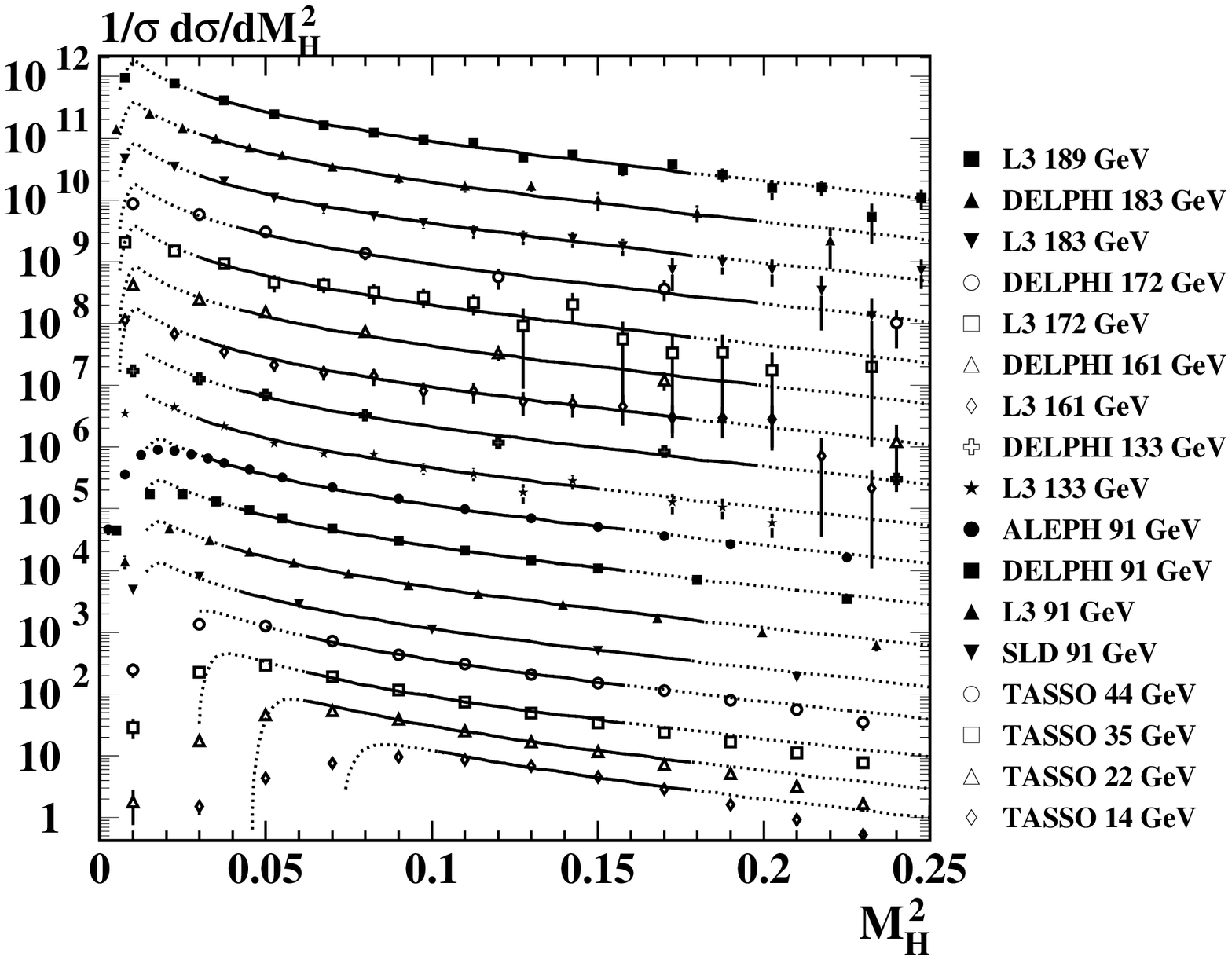}
\caption[bla]{ Scaled distributions for \mh\ and \mhsq\ measured at
$\rs=14$ to 189~GeV. The error bars indicate the total errors of the
data points. The solid lines show the result of the simultaneous fit
of \asmz\ and \azero\ using resummed \oaa+NLLA QCD predictions with
the ln(R)-matching combined with power corrections. The dotted lines
represent an extrapolation of the fit result. }
\label{fig-mh_plot}
\end{center}
\end{figure}

\begin{figure}[!htb]
\begin{center}
\begin{tabular}{cc}
\includegraphics[height=0.6\textheight]{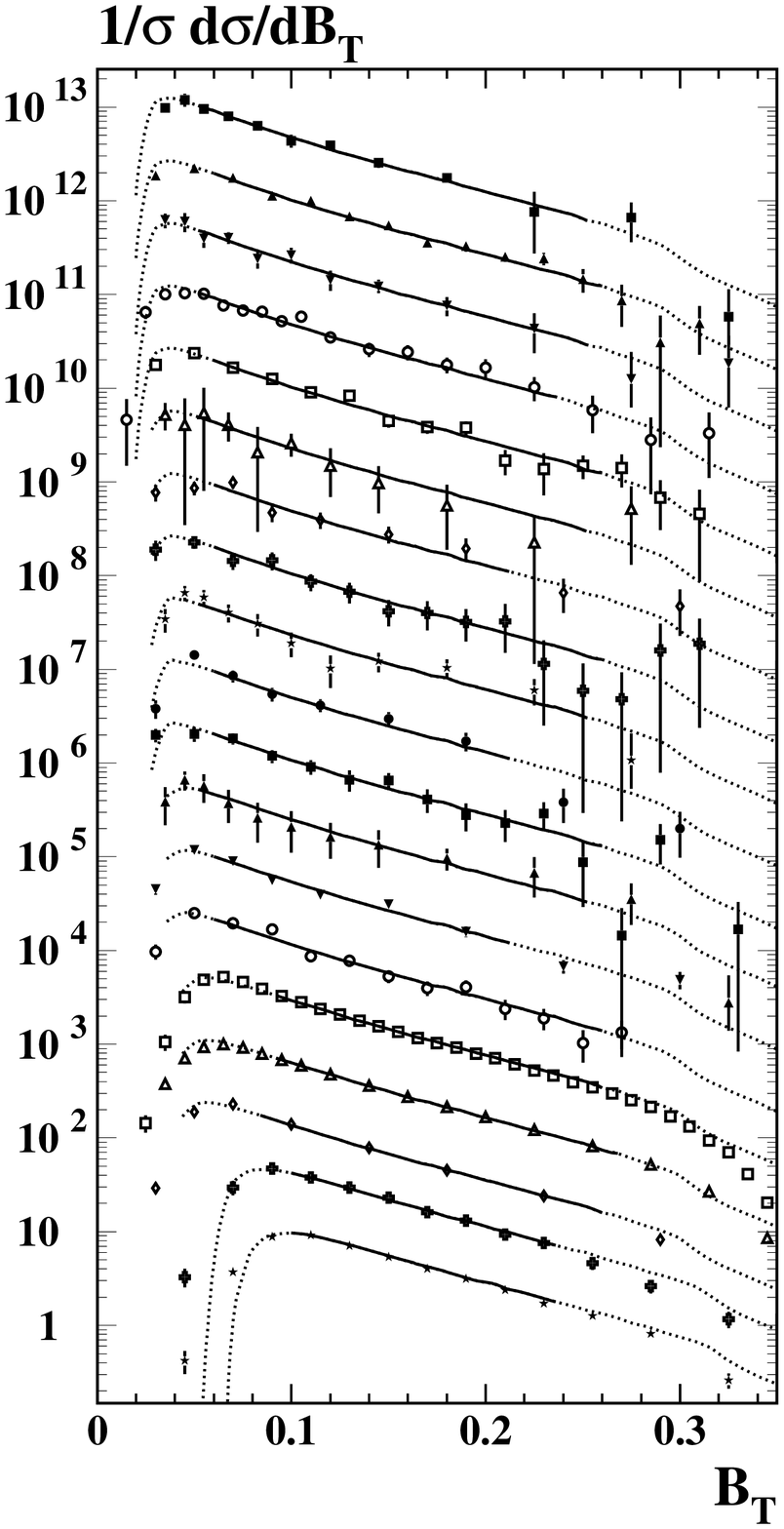} &
\includegraphics[height=0.6\textheight]{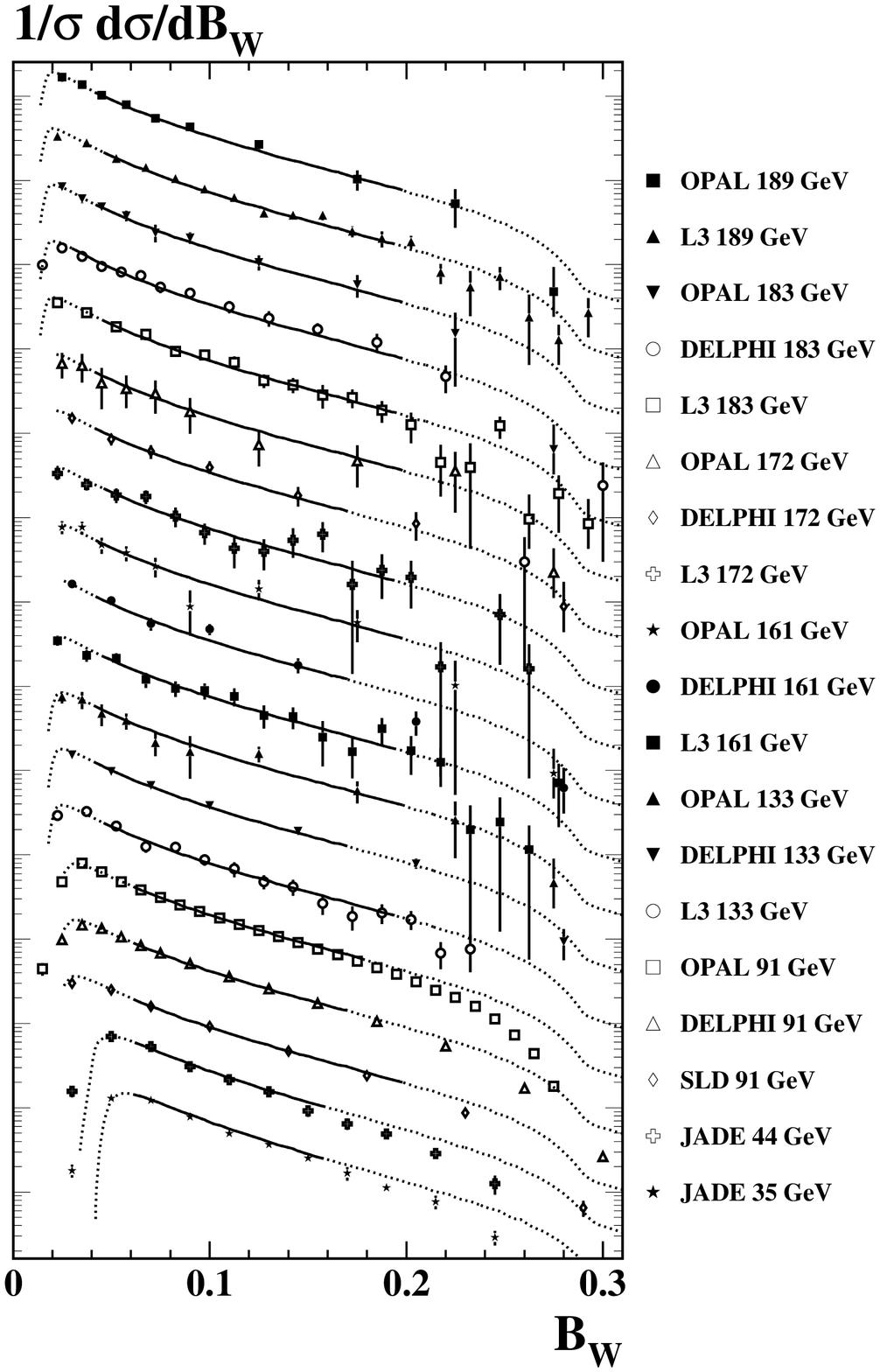}
\end{tabular}
\caption[bla]{ Scaled distributions for \bt\ and \bw\ measured at
$\rs=35$ to 189~GeV. The error bars indicate the total errors of the
data points. The solid lines show the result of the simultaneous fit
of \asmz\ and \azero\ using resummed \oaa+NLLA QCD predictions with
the ln(R)-matching combined with power corrections. The dotted lines
represent an extrapolation of the fit result. }
\label{fig-bt_plot}
\end{center}
\end{figure}

\begin{figure}[!htb]
\begin{center}
\includegraphics[width=0.7\textwidth]{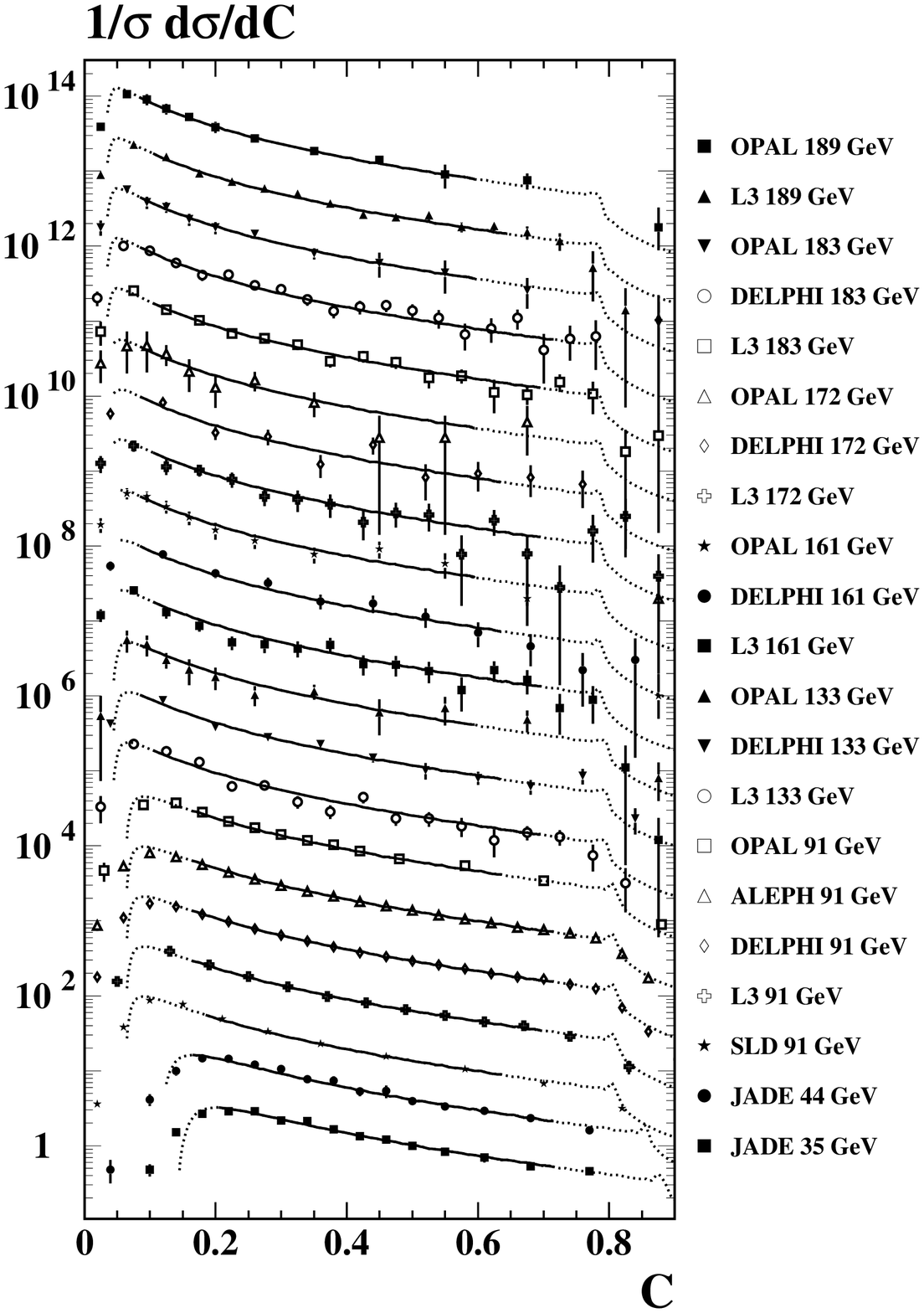}
\caption[bla]{ Scaled distributions for \cp\ measured at
$\rs=35$ to 189~GeV. The error bars indicate the total errors of the
data points. The solid lines show the result of the simultaneous fit
of \asmz\ and \azero\ using resummed \oaa+NLLA QCD predictions with
the ln(R)-matching combined with power corrections. The dotted lines
represent an extrapolation of the fit result. }
\label{fig-c_plot}
\end{center}
\end{figure}

Our standard results for an observable are obtained with $\xmu=1$ and
$\mui=2$~GeV. The results from the fits are listed in
tables~\ref{tab-as-powcor} and~\ref{tab-a0-powcor} for
\asmz\ and \azero, respectively. The signed values indicate the
direction in which \asmz\ and \azerotwo\ changed w.r.t. the standard
analysis when systematic effects are studied.  The fit curves of the
standard results and the corresponding experimental data for \thr,
\mh\ or \mhsq, \cp, \bt\ and \bw\ are shown in figures~\ref{fig-t_plot}
to~\ref{fig-c_plot}. Values of \chisqd\ and of the correlation
coefficients from the fits are given in table~\ref{tab_chisum}.

We generally observe a good agreement of the predictions with the data
within the fit ranges, as indicated by the values of \chisqd 
The two fit parameters are anticorrelated with a correlation
coefficient $\rhofit\simeq -80\perc$.  The results for \asmz\ and
\azerotwo\ are consistent with each other within the total errors in
the case of \thr, \cp\ and \bt. 

The agreement between data for \bw\ at $\rs<\mz$ and the QCD
prediction is not as good as with the other observables, see also
table~\ref{tab_chisum}.  The value for \asmz\ obtained for
\bw\ is about 15\perc\ smaller than the values from the other
observables. In the case of \bw\ the QCD prediction in the 3-jet
regions tends to lie above the data leading to smaller values of
\asmz\ in the fit. Fitting only data at $\rs<\mz$ leads to a
significant deviation of \asmz\ w.r.t. the standard result, see
table~\ref{tab-as-powcor}. This may indicate that \rs-dependent
non-perturbative effects are not fully modelled by the calculations
for \bw. 

In order to disentangle the different contributions to this effect we
performed a fit of the \oaa\ QCD prediction for \bw\ combined with power
corrections with \asmz, \xmu\ and \azerotwo\ as free parameters and
obtained $\asmz=0.106\pm0.001$, $\xmu=0.10\pm0.02$ and
$\azerotwo=0.65\pm0.03$ with $\chisqd=0.5$. Since the value for \asmz\
is comparatively small and the value for \azerotwo\ is comparatively
large we conclude that both the \oaa+NLLA perturbative predictions
and the power correction calculations contribute to the small values
of \asmz\ and large values of \azerotwo\ observed in the standard fits. 
Small values of \asmz\ in fits with \bw\
using \oaa+NLLA QCD calculations have also been observed
in~\cite{jadenewas,jadec,OPALPR075,delphinlla,sldnlla,alephnlla,l3nlla}.

The results for \azerotwo\ are consistent with each other within about
two standard deviations of the total errors; in particular the values
for \azerotwo\ from \mh\ or \mhsq\ and \bw\ are approximately 25\perc\
larger than the other results. We note the coincidence that \mh\ and
\bw\ are calculated using only the hemispheres containing more invariant
mass or transverse momentum, respectively.  We conclude that
\azerotwo\ is approxiately universal within the total uncertainties of
the individual measurements. The results for \azerotwo\ are also
consistent with earlier
measurements~\cite{jadec,l3lep2data1,delphilep2data1}.

The values of \asmz\ obtained from the fits are
systematically lower than corresponding results which use the same
\oaa+NLLA perturbative predictions but apply Monte Carlo 
corrections instead of power corrections
~\cite{jadenewas,jadec,OPALPR075,delphinlla,sldnlla,alephnlla,l3nlla}.
From the experimental point of view there is a lucid explanation for
the differences between the \asmz\ results based on the power
corrections and those based on Monte Carlo
corrections~\cite{movilla98}. The latter induce a stronger squeeze to
all distributions than the power corrections which simply predict a
shift for \thr, \mhsq\ and \cp\ without any presence of a squeeze.
Although the situation improved for the jet broadening observables
due to the revised calculations~\cite{dokshitzer99a} the effect of the
squeeze remains below the expectation of the Monte Carlo hadronisation
models.  As a consequence the two-parameter fit favours smaller values
for \asmz\ in order to make the predicted shape more peaked in the 2-jet
region and hence chooses large values for
\azero\ in order to compensate the shift of the distribution towards
the 2-jet region. 

Figure~\ref{fig_pcr} compares hadronisation corrections as predicted by
power corrections and by the JETSET Monte Carlo program as used in
section~\ref{sec_bcorr}. The hadronisation corrections from the Monte
Carlo simulation are defined as the ratio of distributions calculated
using the partons left at the end of the parton shower (parton-level)
and the stable particles ($\tau>300$~ps) after hadronisation and
decays (hadron-level). In all cases and in particular for \mh\ and
\bw\ the Monte Carlo corrections increase the slopes of the
perturbative predictions more than the power corrections leading to
larger values of \asmz\ in fits of the predictions to the data.

It turns out that the power correction uncertainties for \asmz\ are
negligible for each observable while there are significant power
correction uncertainties for \azero. We conclude that \asmz\ is mainly
constrained by the perturbative prediction rather than by the power
correction contributions while \azero\ is mostly determined by the
power correction calculations. The strong dependence of \azero\ on \mil\
is due to the anticorrelation seen in equation~(\ref{equ_pcp}). 

The total errors are generally dominated by the theoretical
uncertainties. We observe significant variations of \asmz\ from \bt\
and \bw\ when considering only data with $\rs<\mz$ in the fits; for
\bw\ this variation is the largest contribution to the total error of
\asmz. For the jet broadening observables the matching scheme
uncertainty is larger than the renormalisation scale uncertainty for
\azerotwo\ and also for \asmz\ in the case of \bt. We also notice that
the \azerotwo\ results from \bw\ and \mh\ or \mhsq\ have the largest
power correction uncertainties.

\subsection{ Results of Fits to Mean Values }

\begin{figure}
\begin{center}
\begin{tabular}{cc}
\includegraphics[width=0.4\textwidth]{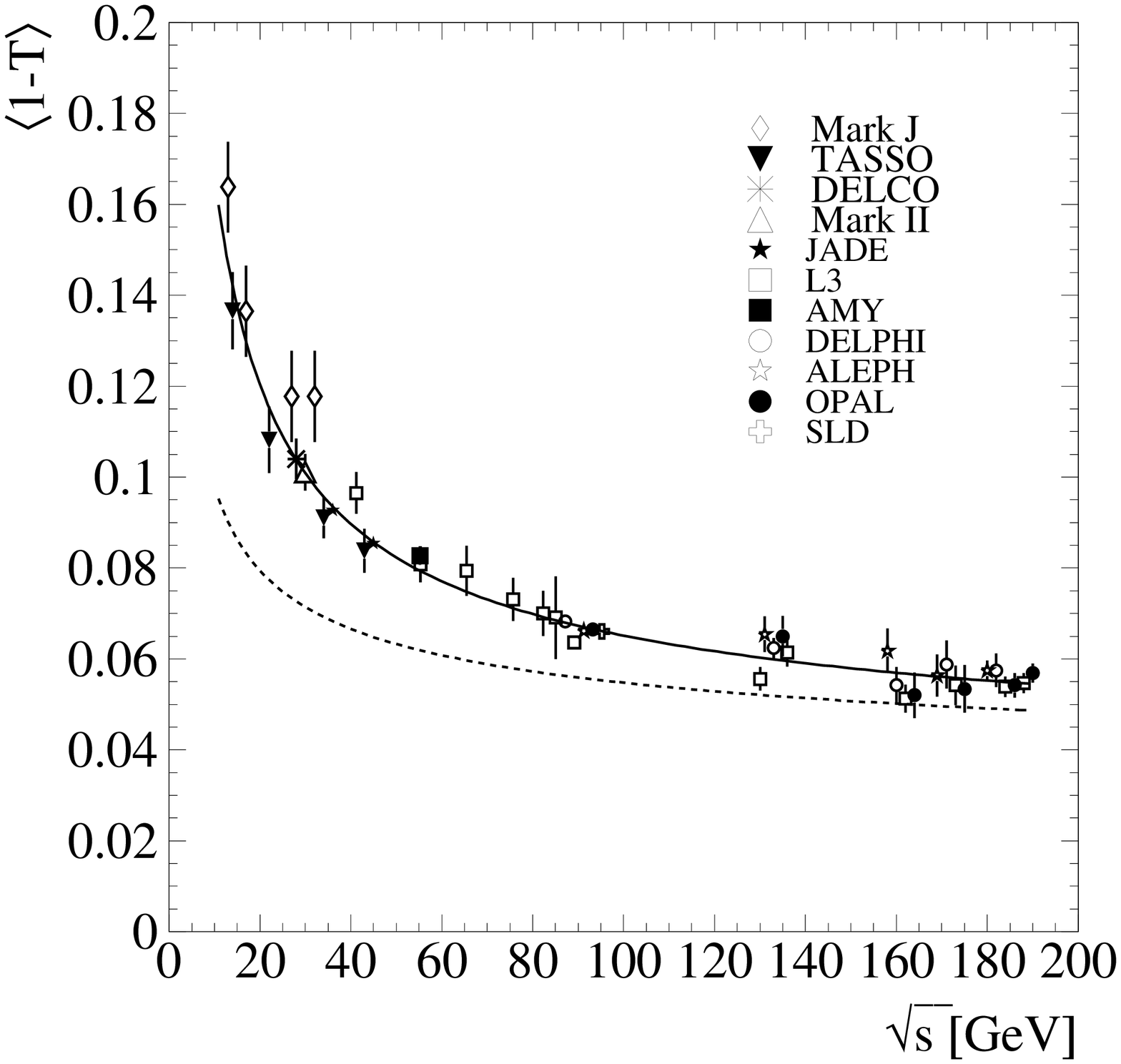} &
\includegraphics[width=0.4\textwidth]{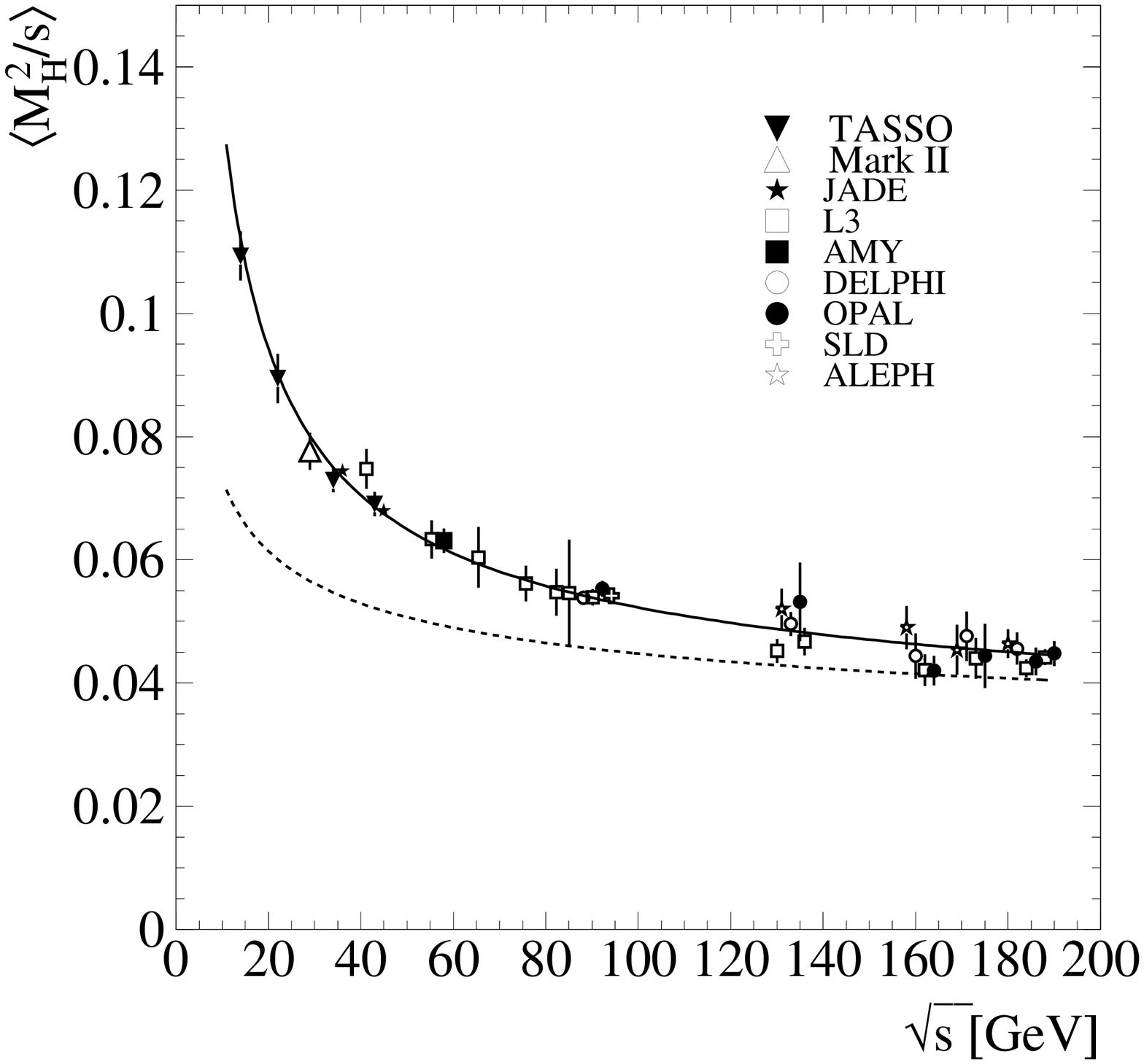} \\
\includegraphics[width=0.4\textwidth]{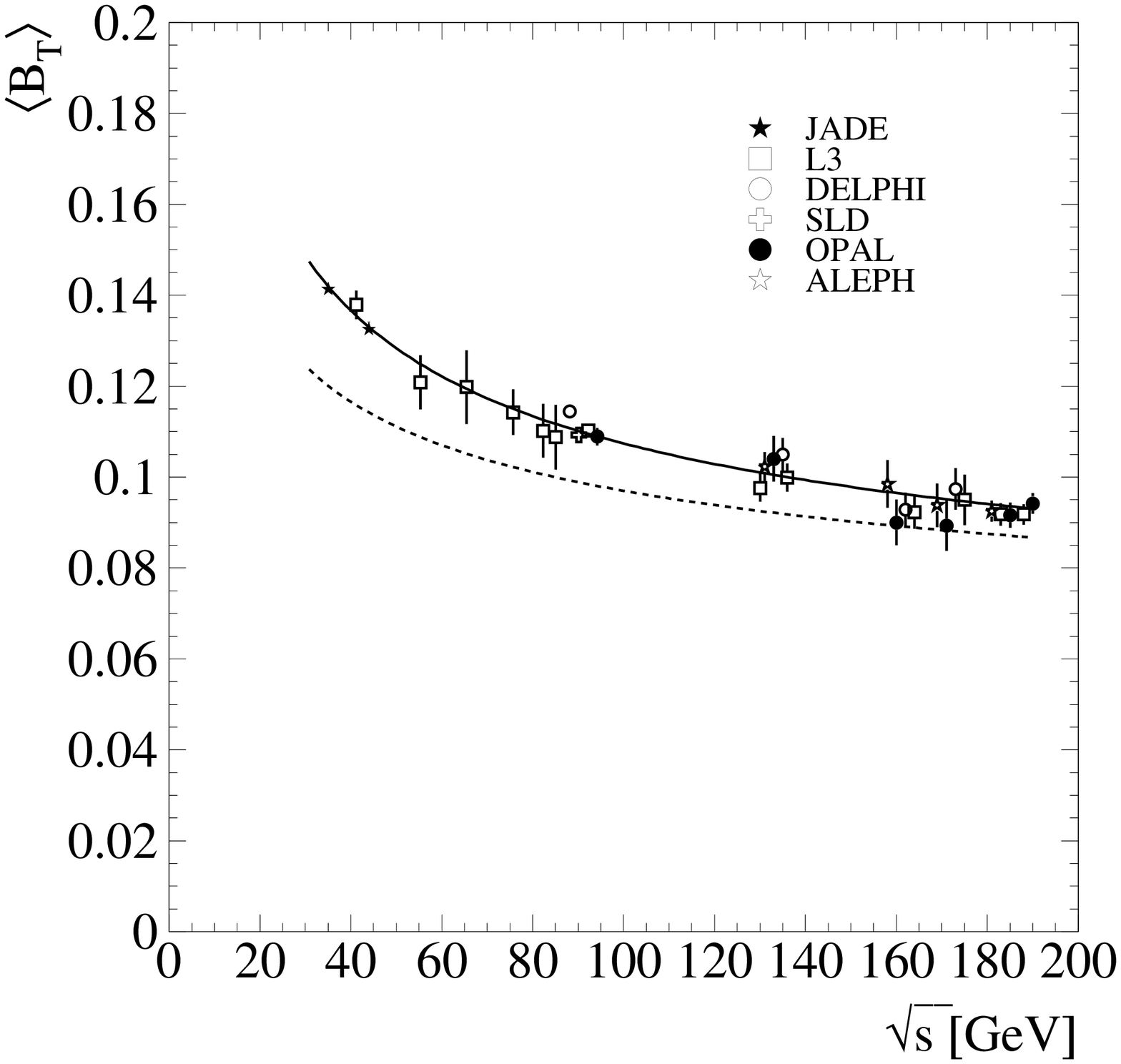} &
\includegraphics[width=0.4\textwidth]{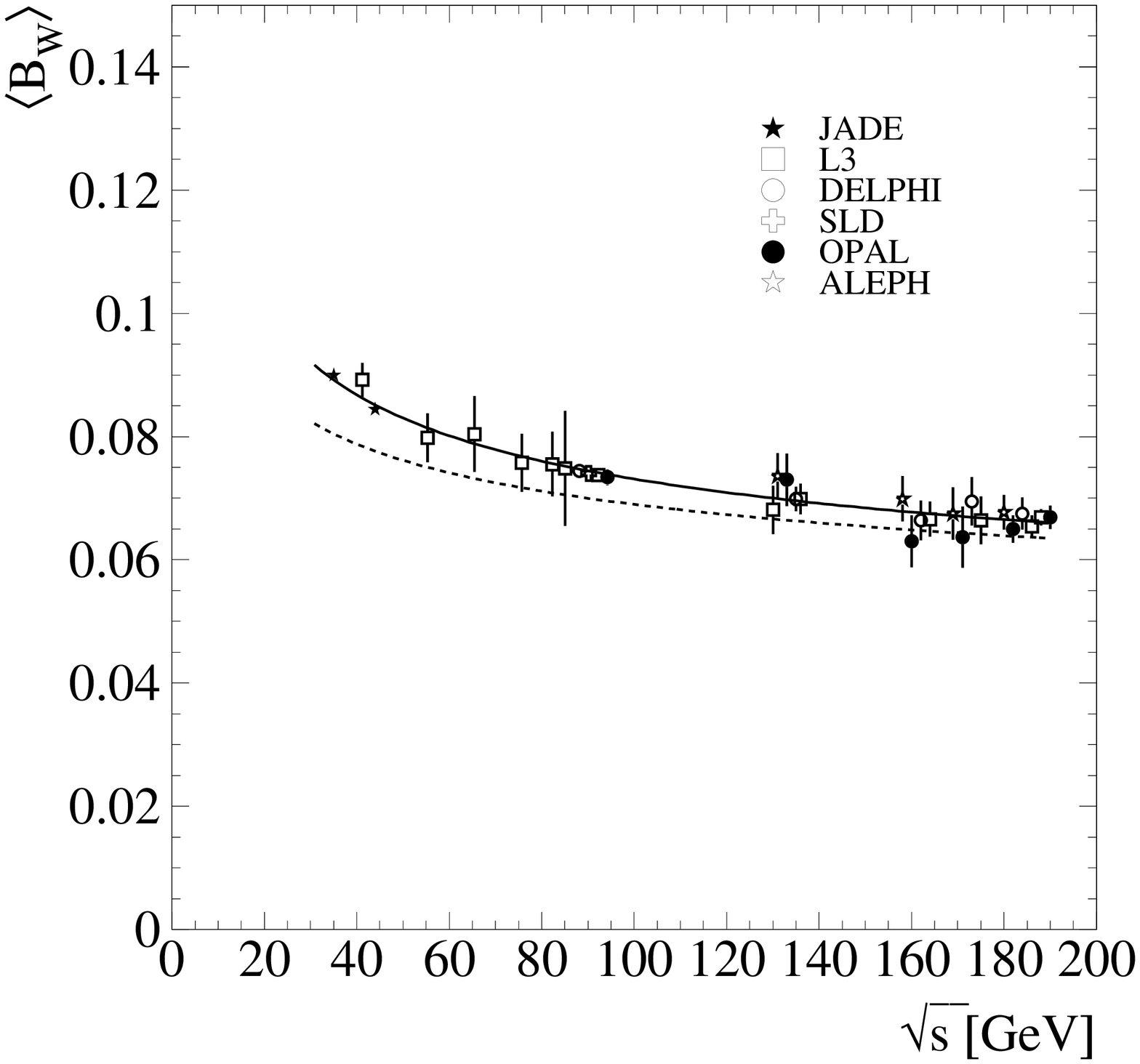} \\
\includegraphics[width=0.4\textwidth]{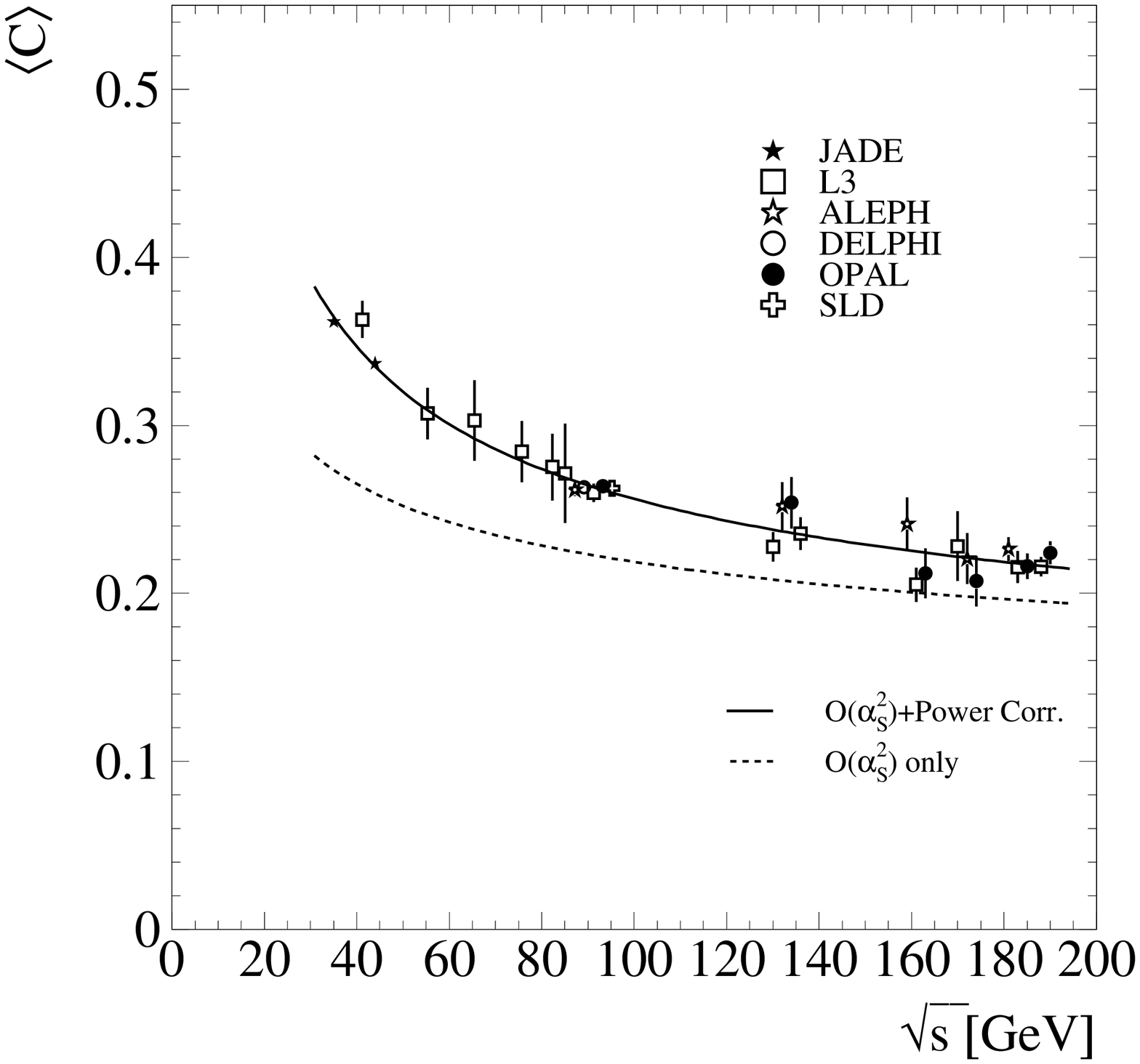} &
\includegraphics[width=0.4\textwidth]{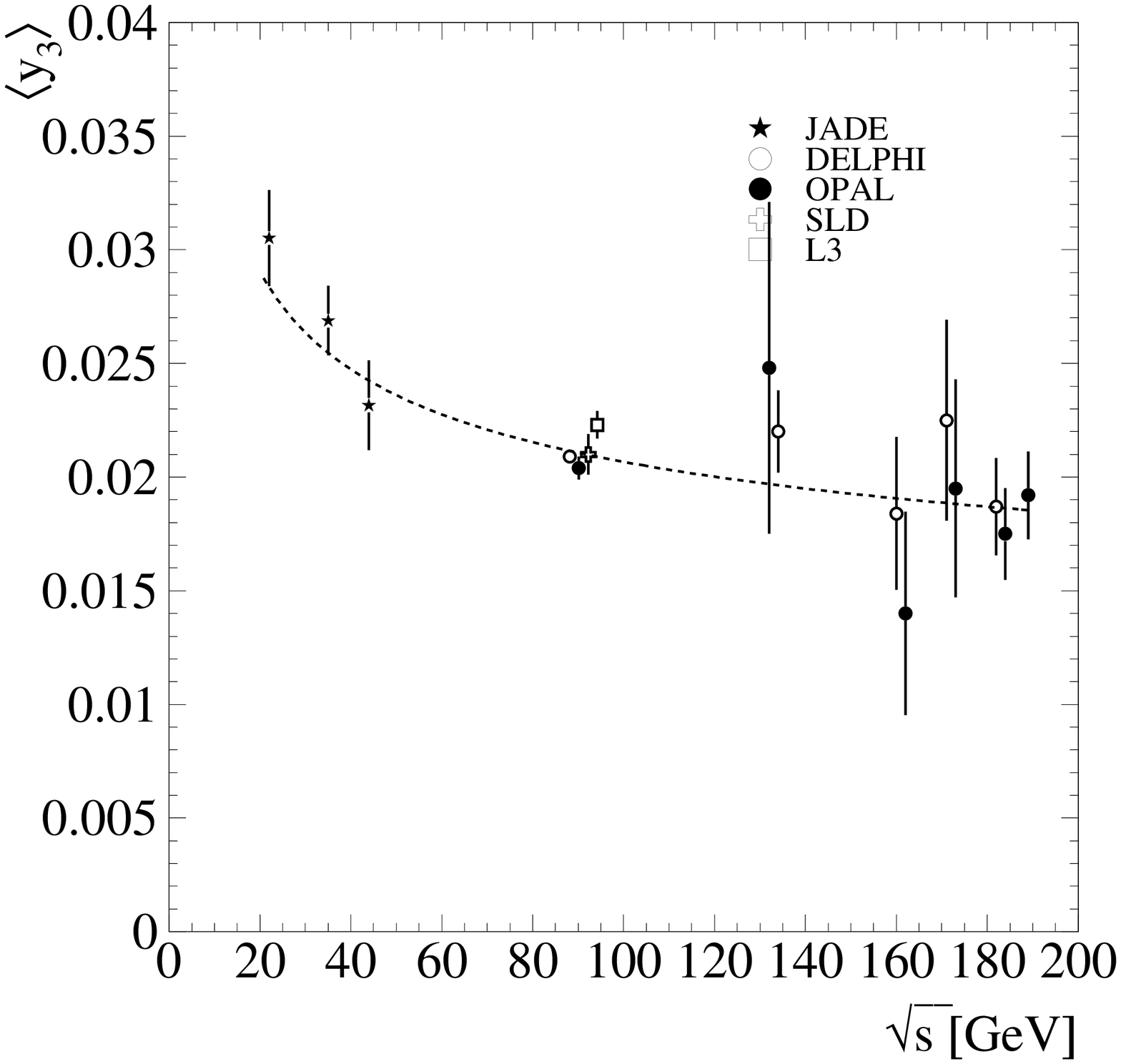} \\
\end{tabular}
\caption[bla]{ The energy dependence of \momone{\thr}, \momone{\mhsq},
\momone{\bt}, \momone{\bw}, \momone{\cp} and \momone{\ythree} is
shown. The solid curves are the results of fits using perturbative
\oaa\ QCD calculations combined with power corrections while the
dashed lines indicate the contribution from the perturbative
prediction only.  }
\label{fig_means} 
\end{center}
\end{figure}

The main fits to the mean values of \thr, \mhsq, \bt, \bw\ and \cp\
are performed with \asmz\ and \azerotwo\ as free parameters using
$\xmu=1$ and $\mui=2$~GeV.  In figure~\ref{fig_means} the results of
the fits and the corresponding perturbative contribution
$\momone{y}_{\mathrm{PT}}$ of equation~(\ref{equ_pcmom}) are
shown. The size of the power suppressed contribution is the difference
between the dashed and the solid curves in
figure~\ref{fig_means}. Tables~\ref{tab_mean_as} and~\ref{tab_mean_a0}
list the results of the fits and the variations found from the studies
of systematic uncertainties.

We find that the fitted QCD predictions describe the data well with
$\chisqd\simeq 1$. The fit results for \asmz\ and \azerotwo\ for all
observables are consistent with each other within their total errors
and have correlation coefficients $\rhofit\simeq -90\perc$. The
results are also generally consistent with the results from fits to
distributions. We note that in contrast to the analysis of
distributions the results for \asmz\ and \azerotwo\ from \mhsq\ and
\bw\ are compatible with results from the other observables. 

For the observable \momone{\ythree} we investigated power corrections
of the form $1/Q^2$, $(\ln Q)/Q^2$, $1/Q$, $(\ln Q)/Q$ and omitting
power correction terms, introducing
$\alpha_1(\mui)=(1/\mui)^2\cdot\int_0^{\mui}k\cdot\as(k)\dd k$ as the
second and an unknown coefficient $D_{\ythree}$ as the third fit
parameter~\cite{biebel01a,movilla00a}. All fits yielded $\chisqd\simeq
1$. For the $1/Q$ and $(\ln Q)/Q$ corrections large values for \asmz\
were obtained which are incompatible with the world
average~\cite{bethke00a} within the fit errors. Corrections of the
$1/Q^2$ and $(\ln Q)/Q^2$ type gave 1 to 2\perc\ increased values of
\asmz\ and a value of $\alpha_1(2\mathrm{GeV})=0.25\pm0.03\fit$. The
results for $D_{\ythree}$ were $-0.2$ and $-0.4$ for $1/Q^2$ and $(\ln
Q)/Q^2$, respectively, but also consistent with zero within the fit
errors.  We conclude that the data prefer one of these latter types of
power corrections although the size of the correction is too small to
be determined from the available data. The smallness of the fitted
$D_{\ythree}$ coefficient justifies to neglect any power correction
for fits of the \momone{\ythree} data and we only quote the result of
such fits in table~\ref{tab_mean_as}. The result for \asmz\ from
\momone{\ythree} is also in good agreement with the world average
value of the strong coupling.

\subsection{ Combination of Individual Results }
\label{sec_comb}

The individual results are combined to single values for \asmz\ and
\azerotwo, respectively, following the procedure described
in~\cite{jadenewas,OPALPR075}. The combination is done separately for
the results from event shape distributions or mean values. A weighted
average of the individual results is calculated with the square of the
reciprocal total errors used as the weights. For each of the systematic
checks the weighted averages for \asmz\ and \azerotwo\ are also
determined and the total error of the weighted average is calculated
exactly as described in section~\ref{sec_syst}. This procedure
accounts for correlations of the systematic errors.

We obtain as combined results from the analysis of distributions
\begin{eqnarray} \nonumber
 \asmz & = & 0.1111\pm0.0004\fit\pm0.0020\syst^{+0.0044}_{-0.0031}\theo
\\ \nonumber
 \azerotwo & = & 0.579\pm0.005\fit\pm0.011\syst^{+0.099}_{-0.071}\theo
\;\;\;.\\ \nonumber
\end{eqnarray}
The error contributions refer to the fit error (fit), the variations
of the input data sets and the fit ranges (syst.) and the variations
of the matching scheme, renormalisation scale, Milan factor and
\epsb\ (theo.). The total correlation coefficient is estimated as
$\rho=-0.16$ (see below). The small value for \asmz\ compared to the
world average $\asmz=0.1181\pm0.0034$~\cite{bethke00a} is caused by
the small values of \asmz\ from \mh\ or \mhsq\ and in particular
\bw. If the results from \bw\ are omitted from the weighted averages
the results become $\asmz=0.1126^{+0.0050}_{-0.0038}$ and 
$\azerotwo=0.558^{+0.093}_{-0.067}$. This value for \asmz\ is in
better agreement with the world average and with other
measurements~\cite{jadec,OPALPR075,delphinlla,sldnlla,alephnlla,l3nlla}
while the result for \azerotwo\ changes only slightly.

The results from the study of mean values based on \momone{\thr},
\momone{\mhsq}, \momone{\bt}, \momone{\bw} and \momone{\cp} are
\begin{eqnarray} \nonumber
 \asmz & = & 0.1187\pm0.0014\fit\pm0.0001\syst^{+0.0028}_{-0.0015}\theo
 \\ \nonumber 
 \azerotwo & = & 0.485\pm0.013\fit\pm0.001\syst^{+0.065}_{-0.043}\theo
  \;\;\;.\\ \nonumber
\end{eqnarray}
The error contributions are defined as explained above for
distributions. The estimate of the total correlation coefficient is
$\rho=+0.17$. The values for \asmz\ and \azerotwo\ are in reasonable
agreement with the results from distributions, especially when the
average of results from distributions is calculated without the values
from \bw.

\begin{figure}[!htb]
\begin{center}
\begin{tabular}{cc}
\includegraphics[width=0.475\textwidth]{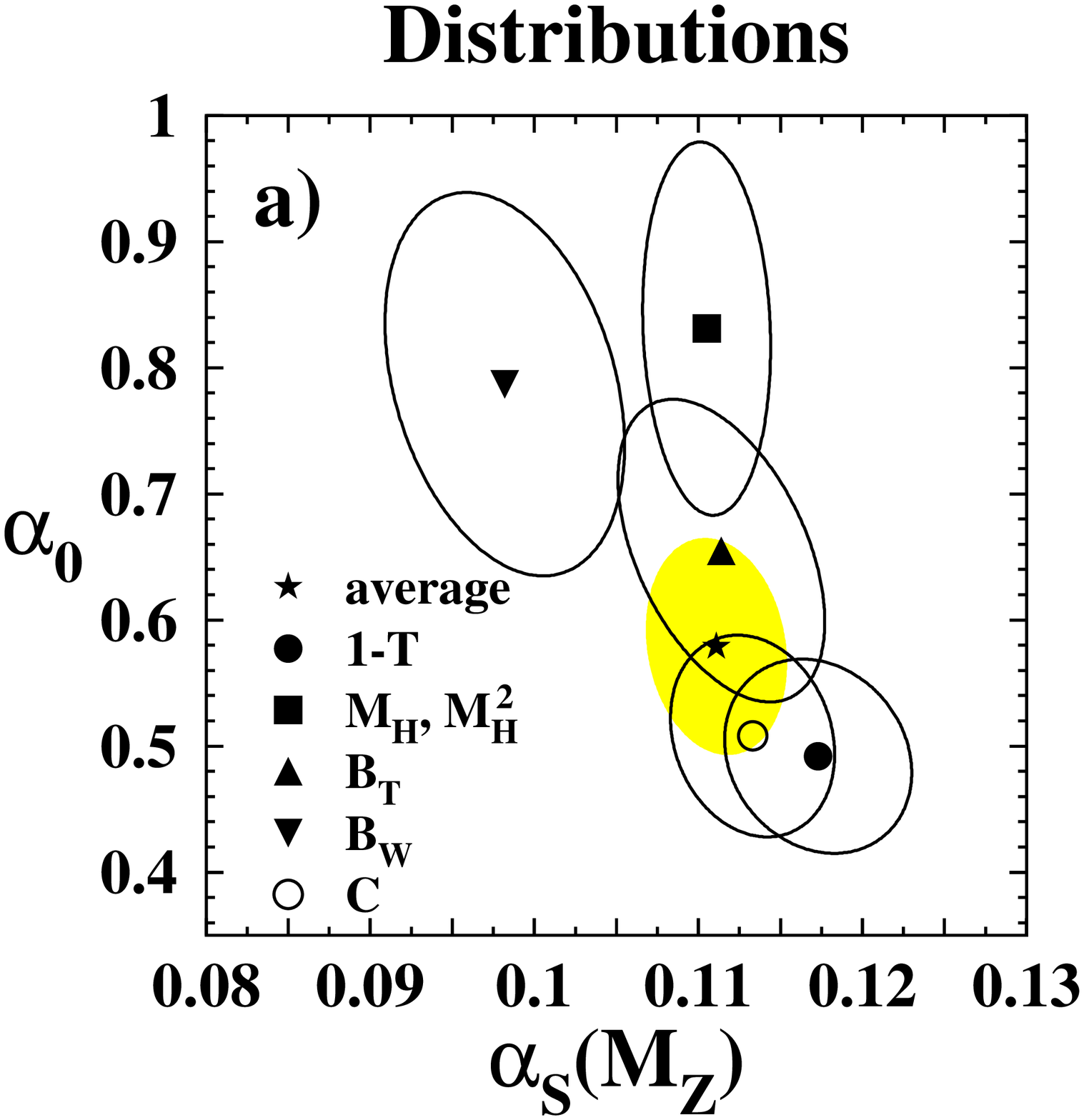} &
\includegraphics[width=0.475\textwidth]{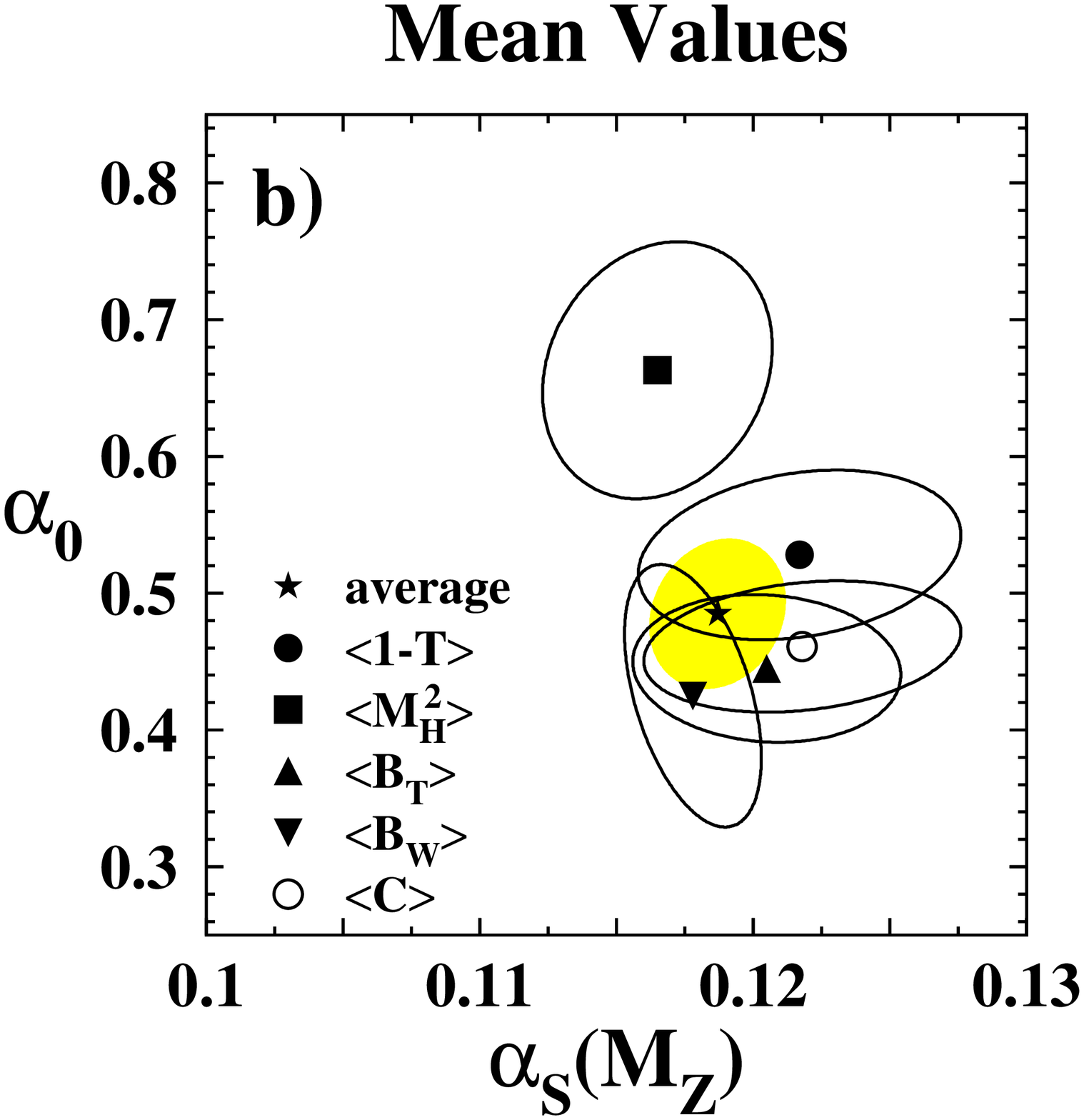} 
\end{tabular}
\caption[bla]{ Results for \asmz\ and \azerotwo\ from fits perturbative
QCD predictions combined with power corrections to 
distributions a) or mean values b) of the event
shape observables  \thr, \mh\ or \mhsq, \bt\ and \bw\ and \cp\ are
shown. The error ellipses correspond to one standard deviation of the
total error (38\perc\ CL) and take correlations from the fit and from
systematic uncertainties into account as explained in the text.}
\label{fig-a0_results}
\end{center}
\end{figure}

Figures~\ref{fig-a0_results} a) and b) present the results for \asmz\
and \azerotwo\ with error ellipses based on the total errors. The
correlation coefficients are determined as follows. For every
systematic test a covariance matrix is constructed using the
systematic uncertainties, symmetrised if neccessary, and a correlation
coefficient \rhosyst. In cases where the correlation between
systematic deviations of \asmz\ and \azerotwo\ of a given systematic
test has the same sign as the correlation \rhofit\ from the fit result
we set $\rhosyst=\rhofit$. In cases where the signs from the
correlations from the standard fit and the systematic test are
opposite we set $\rhosyst=+1$ or $-1$ taking the sign from the
correlation of the systematic test. All covariance matrices are added
and the result defines the error ellipsis. Tables~\ref{tab_chisum}
and~\ref{tab_mean_as} show the correlation coefficients obtained with
this procedure.  The correlation coefficients of the averages are
calculated as the weighted averages of the individual correlation
cofficients using the products of the individual total errors for
\asmz\ and \azerotwo\ as weights. The figure illustrates that the
individual results for \asmz\ and \azerotwo\ from distributions and
mean values are compatible with each other and with the averages
within the total errors. We consider this as a confirmation of the
predicted universality of the non-perturbative parameter \azero.

Finally we combine the results for \asmz\ from the
analysis of distributions, mean values of \thr, \mhsq, \bt, \bw\ and
\cp\ and from \momone{\ythree} by calculating error 
weighted averages based on the symmetrised total errors.  As errors of
the final combined results the smaller of the errors of the individual
results are chosen. The final result for \asmz\ is
\begin{displaymath}
 \asmz = 0.1171^{+0.0032}_{-0.0020} \;\;\;.
\end{displaymath}
The final result for \azerotwo\ is obtained by combining the results
from distributions and mean values again quoting the smaller of the
total errors of the individual results as the final errors:
\begin{displaymath}
  \azerotwo = 0.513^{+0.066}_{-0.045} \;\;\;.\
\end{displaymath}
The total correlation coefficient is again estimated as a weighted
average of the correlation coefficients from the combined results
from distributions and mean values, respectively, yielding $\rho=+0.07$.

\section{ Summary and Conclusions }
\label{sec_conc}

The analytic treatment of non-perturbative effects to event shape
observables in \epem\ annihilation based on power corrections was
examined. We tested predictions for the differential distributions and
mean values of the observables \thr, \mh\ or \mhsq, \bt, \bw, \cp\ and
\ythree, respectively. For this test a large amount of event shape data
collected by several experiments over a range of
\epem\ annihilation energies from $\rs=14$ to 189~GeV was considered.

Fits of perturbative QCD predictions
combined with power corrections to distributions and mean values of
event shape observables were performed with
the strong coupling \asmz\ and the non-perturbative parameter
\azeromui\ as free parameters.  The good quality of the fits with
$\chisqd\simeq1$ supports the predicted $1/Q$ evolution of the power
corrections.  The results for \asmz\ and \azerotwo\ from distributions
are more consistent with each other compared to previous
studies~\cite{movilla98,wicke98} due to the improved predictions of
power corrections to the jet broadening variables.  However, we still
observe a large deviation of the \asmz\ results obtained from \bw\
from those extracted from the other observables. We conjecture that
this discrepancy is a combined effect of the perturbative \oaa+NLLA
predictions and the power correction calculations for this observable.
The individual results for \asmz\ from all observables are observed to
be systematically smaller than the corresponding results
in~\cite{jadec,OPALPR075,delphinlla,sldnlla,alephnlla,l3nlla}, which
use Monte Carlo hadronisation models.  This observation may be related
to the different amounts of squeeze of the distributions predicted by
both types of hadronisation model.

We obtain as combined results for the strong coupling constant 
and the non-perturbative parameter:
\begin{eqnarray} \nonumber
 \asmz & = & 0.1171^{+0.0032}_{-0.0020} \\ \nonumber
 \azerotwo & = & 0.513^{+0.066}_{-0.045} \;\;\;.\\ \nonumber
\end{eqnarray}
It should be noted that the values for \asmz\ and \azerotwo\ from \bw\
are only compatible with the combined result and with the values of
\asmz\ from the other observables within about two standard deviations
of the total errors. 

The average value for \azerotwo\ is in good agreement with 
previous results~\cite{jadec,kluth00a,biebel01a}. The scatter of
\azerotwo\ values derived from \thr, \bt\ and \cp\ is covered by the
expected theoretical uncertainty of the Milan factor of about
20\perc~\cite{dokshitzer98b}. 

Since this value is
representative of the individual results within the errors we consider
this as a confirmation of the universality of \azero\ as predicted by
the DMW model.  However, the results from \mh\ or \mhsq\ and \bw\
from distributions indicate that uncalculated higher orders may
contribute significantly to the non-perturbative corrections.

\clearpage
\section*{ Tables }

\begin{table}[!htb]
\begin{center}
\begin{tabular}{|r|l||c|c|c|c|c|}
\hline
\rs    & Experiment & \thr & \mh, \mhsq & \bt & \bw & \cp \\ \hline\hline
189    & L3 \cite{l3lep2data1} & $0.025-0.30$ & $0.03-0.18$ & 
$0.06-0.26$ & $0.045-0.195$ & $0.10-0.65$ \\
       & OPAL \cite{OPALPR303} & $0.03-0.30$ & $0.14-0.45$ & 
$0.05-0.25$ & $0.04-0.20$ & $0.08-0.60$ \\ \hline
183    & DELPHI \cite{delphilep2data1} & $0.03-0.28$ & $0.03-0.20$ & 
$0.05-0.24$ & $0.03-0.20$ & $0.08-0.72$ \\ 
       & L3 \cite{l3lep2data1} & $0.025-0.30$ & $0.03-0.18$ & 
$0.06-0.26$ & $0.045-0.195$&$  0.10-0.70$ \\
       & OPAL \cite{OPALPR303} & $0.03-0.30$ & $0.14-0.45$ & 
$0.05-0.25$ & $0.04-0.20$ & $0.08-0.60$ \\ \hline
172    & DELPHI \cite{delphilep2data1} & $0.04-0.32$ & 
$0.04-0.20$ & $0.06-0.21$ & $0.04-0.17$ & $0.08-0.64$ \\ 
       & L3 \cite{l3lep2data1} & $0.025-0.30$ & $0.03-0.18$ & 
$0.06-0.26$ & $0.045-0.195$&$  0.10-0.70$ \\
       & OPAL \cite{OPALPR303} & $0.03-0.30$ & $0.14-0.45$ & 
$0.05-0.25$ & $0.04-0.20$ & $0.08-0.60$ \\ \hline
161    & DELPHI \cite{delphilep2data1} & $0.04-0.32$ & $0.04-0.20$ & 
$0.0 6-0.21$ & $0.04-0.17$ & $0.08-0.64$ \\
       & L3 \cite{l3lep2data1} & $0.05-0.30$ & $0.03-0.18$ & 
$0.06-0.26$ & $0.045-0.195$& $0.10-0.70$ \\ 
       & OPAL \cite{OPALPR197} & $0.03-0.30$ & $0.14-0.45$ & 
$0.05-0.25$ & $0.04-0.20$ & $0.08-0.60$ \\ \hline
133    & ALEPH \cite{alephas133} & $0.04-0.30$ & & & & \\ 
       & DELPHI \cite{delphilep2data1} & $0.04-0.32$ & $0.04-0.20$ & 
$0.06-0.21$ & $0.04-0.17$ & $0.08-0.64$ \\
       & L3 \cite{l3lep2data1} & $0.05-0.25$ & $0.03-0.15$ & 
$0.06-0.26$ & $0.045-0.195$& $0.10-0.70$ \\
       & OPAL \cite{OPALPR158} & $0.03-0.30$ & $0.14-0.45$ & 
$0.05-0.25$ & $0.04-0.20$ & $0.08-0.60$ \\ \hline
91     & ALEPH \cite{alephlep1datanew} & $0.06-0.30$ & $0.035-0.16$ & & &
$0.16-0.72$ \\
       & DELPHI \cite{delphilep1data} & $0.06-0.30$ & $0.04-0.16$ & 
$0.09-0.27$ & $0.06-0.17$ & $0.16-0.72$ \\
       & L3 \cite{l3lep1data} & $0.065-0.33$ & $0.039-0.183$ & & & 
$0.16-0.70$ \\ 
       & OPAL \cite{OPALPR075,OPALPR054} & $0.06-0.33$ & $0.20-0.40$ &
$0.09-0.26$ & $0.06-0.18$ & $0.16-0.64$ \\
       & SLD \cite{sldnlla} & $0.06-0.32$ & $0.04-0.18$ & 
$0.08-0.26$ & $0.06-0.20$ & $0.18-0.64$ \\ \hline
55     & AMY \cite{amydata} & $0.10-0.30$ & & & & \\ \hline
44     & JADE \cite{jadenewas,jadec} & $0.06-0.30$ & $0.22-0.42$ & 
$0.10-0.24$ & $0.06-0.16$ & $0.16-0.72$ \\
       & TASSO \cite{tassodata} & $0.06-0.32$ & $0.06-0.16$ & & & \\ \hline
35     & JADE \cite{jadenewas,jadec} & $0.06-0.30$ & $0.22-0.38$ & 
$0.10-0.24$ & $0.06-0.16$ & $0.20-0.72$ \\
       & TASSO \cite{tassodata} & $0.06-0.32$ & $0.06-0.16$ & & & \\ \hline
29     & HRS \cite{hrsdata} & $0.10-0.325$ & & & & \\
       & MARKII \cite{markiidata} & $0.10-0.32$ & & & & \\ \hline
22     & TASSO \cite{tassodata} & $0.10-0.32$ & $0.06-0.18$ & & & \\ \hline
14     & TASSO \cite{tassodata} & $0.12-0.32$ & $0.10-0.18$ & & & \\
\hline
\end{tabular}
\caption[bla]{ The sources of the data and the fit ranges 
for the observables \thr, \mh\ or \mhsq, \bt, \bw\ and \cp\ are
shown. The cms energy \rs\ at which the experiments analysed their
data is given in GeV. The observable \mh\ is used only by OPAL and
JADE while \mhsq\ is used by the other experiments. } 
\label{tab_data}
\end{center}
\end{table}


\begin{table}[!htb]
\begin{center} 
\begin{tabular}{|r||c|c|c|c|c|} \hline 
& \bm{\thr} & \bm{\mh, \mhsq} & \bm{\bt} & \bm{\bw} & \bm{\cp} 
 \\ \hline\hline 
 \bm{\asmz} & \bf 0.1173 & \bf 0.1105 & \bf 0.1114 & \bf 0.0982 
& \bf 0.1133 \\ \hline\hline 
 fit error & $\pm0.0005$ & $\pm0.0005$ & $\pm0.0006$ & $\pm0.0005$ 
& $\pm0.0004$  \\ \hline\hline
 mod. \lnr & $ +0.0013$ & $ +0.0005$ & $ +0.0053$ & $+0.0019$ 
& $ +0.0014$ \\ \hline
 mod. R    & $-0.0010$ & $-0.0005$ & $-0.0021$ & $-0.0012$ & $-0.0005$
\\ \hline
$\xmu=0.5$ & $-0.0041$ & $-0.0023$ & $-0.0039$ & $-0.0011$ & $-0.0039$
\\ \hline
$\xmu=2.0$ & $+0.0055$ & $+0.0037$ & $+0.0050$ & $+0.0023$ & $+0.0052$
\\ \hline\hline
$M-20\%$   & $+0.0003$ & $+0.0002$ & $+0.0001$ & $<0.0001$ & $<0.0001$
\\ \hline
$M+20\%$   & $-0.0003$ & $-0.0002$ & $-0.0001$ & $<0.0001$ & $+0.0001$
\\ \hline
$\mui=1$~GeV & $+0.0007$ & $+0.0004$ & $+0.0003$ & $+0.0001$
& $+0.0001$ \\ \hline  
$\mui=3$~GeV & $-0.0006$ & $-0.0004$ & $-0.0002$ & $-0.0001$ 
& $-0.0001$ \\ \hline\hline 
$\epsb\pm 1\sigma$ & $\pm 0.0002$ & $<0.0001$ & $<0.0001$ 
& $\pm0.0001$ & $<0.0001$ \\ \hline\hline 
$\rs\ge\mz$ & $+0.0023$ & $-0.0022$ & $-0.0005$ & $-0.0008$ & $<0.0001$ 
\\ \hline 
$\rs<\mz$ & $-0.0022$ & $-0.0006$ & $-0.0032$ & $-0.0068$ & $-0.0004$ 
\\ \hline 
$\rs\ne\mz$ & $-0.0026$ & $-0.0014$ & $-0.0006$ & $-0.0002$ & $-0.0018$
\\ \hline 
fit range & $\pm0.0013$ & $\pm0.0010$ & $\pm0.0009$ & $\pm0.0014$ 
& $\pm0.0005$ \\ \hline\hline
\raisebox{-1.5ex}[1.5ex]{\bf total error} 
 & \bm{+0.0063} & \bm{+0.0045} & \bm{+0.0063} & \bm{+0.0073} 
 & \bm{+0.0056} \\
 & \bm{-0.0051} & \bm{-0.0034} & \bm{-0.0063} & \bm{-0.0072} 
 & \bm{-0.0044} \\ \hline
\end{tabular}
\caption{ Values of \asmz\ are shown derived from fits of resummed
\oaa+NLLA QCD predictions combined with power corrections to
distributions of the event shape observables \thr, \mh\ or \mhsq, \bt,
\bw\ and \cp. In addition, the statistical and systematic
uncertainties are given. Signed values indicate the direction in which
\asmz\ changed with respect to the standard analysis.  }
\label{tab-as-powcor} 
\end{center}
\end{table}


\begin{table}[!htb]
\begin{center} 
\begin{tabular}{|r||c|c|c|c|c|} 
\hline 
& \bm{\thr} & \bm{\mh,\mhsq} & \bm{\bt} & \bm{\bw} & \bm{\cp} \\ 
\hline\hline 
 \bm{\azerotwo} & \bf 0.492 & \bf 0.831 & \bf 0.655 & \bf 0.787 
 & \bf 0.507 \\ \hline\hline 
fit error & $\pm0.009$ & $\pm0.011$ & $\pm0.010$ & $\pm0.016$ 
 & $\pm0.005$ \\ \hline\hline
mod. \lnr & $-0.013$ & $-0.013$ & $-0.070$ & $-0.064$ & $-0.049$ \\ \hline 
mod. R    & $+0.008$ & $+0.001$ & $+0.019$ & $+0.020$ & $-0.002$ \\ \hline 
$\xmu=0.5$  & $-0.012$ & $-0.021$ & $-0.010$ & $-0.038$ & $-0.015$ \\ \hline 
$\xmu=2.0$  & $+0.009$ & $+0.014$ & $+0.007$ & $+0.027$ & $+0.011$ \\ \hline 
$M-20\%$    & $+0.063$ & $+0.151$ & $+0.113$ & $+0.155$ & $+0.077$ \\ \hline 
$M+20\%$    & $-0.042$ & $-0.101$ & $-0.075$ & $-0.103$ & $-0.051$ \\ 
\hline\hline 
$\epsb\pm 1\sigma$ & $\pm0.003$ & $\pm0.001$ & $<0.001$ 
& $\pm 0.001$ & $<0.001$ \\ \hline\hline 
$\rs\ge\mz$ & $-0.050$ & $+0.073$ & $+0.012$ & $+0.039$ & $<0.001$ \\ \hline 
$\rs<\mz$   & $+0.028$ & $-0.018$ & $-0.005$ & $<0.001$ & $+0.003$ \\ \hline 
$\rs\ne\mz$ & $+0.029$ & $-0.014$ & $-0.010$ & $-0.040$ & $+0.008$ \\ \hline 
fit range & $\pm0.019$ & $\pm0.018$ & $\pm0.018$ & $\pm0.024$ 
 & $\pm0.003$ \\ \hline\hline
\raisebox{-1.5ex}[1.5ex]{\bf total error} 
 & \bm{+0.084} & \bm{+0.170} & \bm{+0.135} & \bm{+0.175} 
 & \bm{+0.092} \\ 
 & \bm{-0.070} & \bm{-0.128} & \bm{-0.105} & \bm{-0.131} 
 & \bm{-0.071} \\ \hline 
\end{tabular}
\caption{ Values of \azero\ are shown derived from fits of resummed
\oaa+NLLA QCD predictions combined with power corrections to
distributions of the event shape observables \thr, \mh\ or \mhsq, \bt,
\bw\ and \cp. In addition, the statistical and systematic
uncertainties are given. Signed values indicate the direction in which
\azero\ changed with respect to the standard analysis.  }
\label{tab-a0-powcor} 
\end{center}
\end{table}

\begin{table}[!htb]
\begin{center} 
\begin{tabular}{|r||r@{/}l|r@{/}l|r@{/}l|r@{/}l|r@{/}l|} \hline  
 & \multicolumn{2}{c|}{\thr} & \multicolumn{2}{c|}{\mh, \mhsq} 
 & \multicolumn{2}{c|}{\bt} & \multicolumn{2}{c|}{\bw} 
 & \multicolumn{2}{c|}{\cp} \\ \hline\hline 
standard fit & 172  &  263  & 137  & 161  & 91.9 
             & 159  &  96.1 & 132  & 150  & 208 \\ \hline
fit correlation 
 & \multicolumn{2}{c|}{$-0.88$} & \multicolumn{2}{c|}{$-0.75$} 
 & \multicolumn{2}{c|}{$-0.85$} & \multicolumn{2}{c|}{$-0.81$} 
 & \multicolumn{2}{c|}{$-0.82$} \\ \hline
total correlation
 & \multicolumn{2}{c|}{$-0.17$} & \multicolumn{2}{c|}{$-0.10$} 
 & \multicolumn{2}{c|}{$-0.47$} & \multicolumn{2}{c|}{$-0.32$} 
 & \multicolumn{2}{c|}{$-0.17$} \\ \hline\hline
$\rs>\mz$ & 73.5 &  131  & 43.3 &  97  & 67.4 
          & 115  &  50.1 & 100  & 95.2 & 134 \\ \hline
$\rs=\mz$ & 43.2 &   59  & 68.0 & 33   & 16.1 
          & 28   &  22.9 &  20  & 44.2 & 49  \\ \hline
$\rs<\mz$ & 55.3 &   69  & 25.4 &  27  & 8.4 
          & 12   &  23.1 &   8 & 10.6 & 21 \\ \hline
\end{tabular}
\caption[bla]{ The values of \chisqd\ and the correlation
coefficients between \asmz\ and \azero\ are shown for the standard fit
in the first two rows. The third row shows the total correlation
coefficients between \asmz\ and \azero\ including effects of
systematic variations of the analysis (see section~\ref{sec_comb} for
details). The other rows present values of \chisqd\ obtained from fits
with subsets of the data with $\rs>\mz$, $\rs=\mz$ or $\rs<\mz$,
respectively. }
\label{tab_chisum}
\end{center}
\end{table}


\begin{table}
\begin{center}
\begin{tabular}{|r||c|c|c|c|c|c|}   \hline
 &\multicolumn{1}{c|}{\bm{\momone{\thr}}}
 &\multicolumn{1}{c|}{\bm{\momone{\mhsq}}}
 &\multicolumn{1}{c|}{\bm{\momone{\bt}}}
 &\multicolumn{1}{c|}{\bm{\momone{\bw}}}
 &\multicolumn{1}{c|}{\bm{\momone{\cp}}}
 &\multicolumn{1}{c|}{\bm{\momone{\ythree}}} \\ \hline\hline
\bm{\asmz}
    &\bf 0.1217  &\bf 0.1165  &\bf 0.1205  &\bf 0.1178  &\bf 0.1218 
&\bf 0.1199  \\ \hline\hline
fit error & $\pm 0.0014$ & $\pm 0.0016$ & $\pm 0.0015$ & $\pm 0.0015$
& $\pm 0.0014$ & $\pm 0.0008$ \\  \hline
\chisqd & $50.1/41$  & $24.0/35$  & $23.7/28$  & $10.4/29$  &
$18.4/26$  & $13.2/15$  \\ \hline
fit corr. & $-0.89$ & $-0.89$ & $-0.91$ & $-0.94$ & $-0.93$ & n.a. 
\\ \hline
total corr. & $0.23$ & $0.18$ & $-0.09$ & $-0.47$ & $0.22$ & n.a. 
\\ \hline\hline
$\xmu=0.5$ &  $-0.0048$ &  $-0.0026$ &  $-0.0037$ &  $+0.0017$ &
$-0.0045$ &  $-0.0039$  \\ \hline
$\xmu=2.0$ &  $+0.0059$ &  $+0.0037$ &  $+0.0048$ &  $+0.0003$ &
$+0.0056$ &  $+0.0050$   \\ \hline
${\cal M}-20\%$ &  $+0.0020$ &  $+0.0011$ &  $+0.0014$ &  $+0.0009$ &
$+0.0020$ & n.a.    \\  \hline
${\cal M}+20\%$ &  $-0.0016$ &  $-0.0009$ &  $-0.0012$ &  $-0.0008$ &
$-0.0016$ & n.a.      \\ \hline
$\mui=1$~GeV &  $+0.0009$ &  $+0.0005$ &  $+0.0006$ &  $+0.0004$ &
$+0.0009$ & n.a.       \\ \hline
$\mui=3$~GeV &  $-0.0009$ &  $-0.0005$ &  $-0.0006$ &  $-0.0004$ &
$-0.0008$ & n.a.     \\ \hline\hline
$\epsb\pm1\sigma$ &  $\pm0.0002$ &  $\pm0.0001$ &  $\pm0.0001$ 
& $\pm0.0001$ & $\pm0.0002$ &  $<0.0001$  \\  \hline\hline

$\rs\ne\mz$ & $+0.0008$ & $-0.0020$ & $-0.0012$ & $+0.0005$ 
 & $+0.0015$ & $+0.0030$ \\ \hline\hline

\raisebox{-1.5ex}[1.5ex]{\bf total error} 
 & \bm{+0.0065} & \bm{+0.0047} & \bm{+0.0054} & \bm{+0.0025} 
 & \bm{+0.0064} & \bm{+0.0059} \\
 & \bm{-0.0054} & \bm{-0.0038} & \bm{-0.0044} & \bm{-0.0025} 
 & \bm{-0.0053} & \bm{-0.0050} \\ \hline
\end{tabular}
\caption{ Values of \asmz\ are shown from fits of \oaa\ QCD
predictions combined with power corrections to mean values of \thr,
\mhsq, \bt, \bw, \cp\ and \ythree.
  Statistical and systematic uncertainties are also
given. Signs indicate the direction in which \asmz\ changes w.r.t. the
standard analysis. The renormalisation and infrared scale
uncertainties are added asymmetrically to the errors of \asmz. } 
\label{tab_mean_as} 
\end{center}
\end{table}


\begin{table}
\begin{center}
\begin{tabular}{|r||c|c|c|c|c|}   \hline
 &\multicolumn{1}{c|}{\bm{\momone{\thr}}}
 &\multicolumn{1}{c|}{\bm{\momone{\mhsq}}}
 &\multicolumn{1}{c|}{\bm{\momone{\bt}}}
 &\multicolumn{1}{c|}{\bm{\momone{\bw}}}
 &\multicolumn{1}{c|}{\bm{\momone{\cp}}} \\ \hline\hline
\bm{\azerotwo}
 &\bf 0.528   &\bf 0.663   &\bf 0.445   &\bf 0.425   &\bf 0.461 \\ \hline
fit error
 &$\pm0.015 $&$\pm0.024 $&$\pm0.020 $&$\pm0.029 $&$\pm0.013 $ \\ \hline\hline
$\xmu=0.5$    
 &  $+0.002 $ &  $+0.010 $ &  $+0.021 $ &  $+0.118 $ &  $+0.004 $ \\ \hline
$\xmu=2.0$     
 &  $-0.001 $ &  $-0.003 $ &  $-0.014 $ &  $-0.046 $ &  $-0.002 $ \\ \hline
${\cal M}-20\%$    
 &  $+0.072 $ &  $+0.107 $ &  $+0.055 $ &  $+0.048 $ &  $+0.054 $ \\ \hline
${\cal M}+20\%$    
 & $-0.049 $ & $-0.072 $ & $-0.037 $ & $-0.032 $ & $-0.037 $ \\ \hline\hline
$\epsb\pm1\sigma$
 &  $\pm0.002 $ & $\pm0.007$ & $\pm0.002$ & $\pm0.002$ & $\pm0.002$ \\
\hline\hline

$\rs\ne\mz$ & $-0.003$ & $+0.017$ & $+0.007$ & $-0.005$ & $-0.006$ \\ 
\hline\hline

\raisebox{-1.5ex}[1.5ex]{\bf total error} 
 & \bm{+0.074} & \bm{+0.111} & \bm{+0.063} & \bm{+0.131} & \bm{+0.056} \\
 & \bm{-0.051} & \bm{-0.078} & \bm{-0.045} & \bm{-0.063} & \bm{-0.040}
\\ \hline
\end{tabular}
\caption{ Values of \azero\ are shown from fits of \oaa\ QCD
predictions combined with power corrections to mean values of \thr,
\mhsq, \bt, \bw\ and \cp.  Statistical and systematic
uncertainties are also given. Signs indicate the direction in which
\azero\ changes w.r.t. the standard analysis. }
\label{tab_mean_a0} 
\end{center}
\end{table}

\clearpage
\section*{ Figures }


\end{document}